\documentclass[acmsmall,screen]{acmart}
\AtBeginDocument{%
  }

\definecolor{darkblue}{rgb}{0, 0, 0.5}
\hypersetup{colorlinks=true, citecolor=darkblue, linkcolor=darkblue, urlcolor=darkblue}

\usepackage{amssymb}
\usepackage{microtype}
\usepackage{url}
\usepackage{booktabs}
\usepackage{amsmath,amsfonts}
\usepackage{algorithmic}
\usepackage{graphicx}
\usepackage{textcomp}
\usepackage{multicol}
\usepackage{multirow}
\usepackage{colortbl}
\usepackage{verbatim}
\usepackage{fancyvrb}
\usepackage{fvextra} 
\usepackage{longtable}
\usepackage{hyperref}
\usepackage{float}

\begin{document}
%% The "title" command has an optional parameter,
%% allowing the author to define a "short title" to be used in page headers.
\title{When AI Meets Finance (StockAgent): Large Language Model-based Stock Trading in Simulated Real-world Environments}

\author{Chong Zhang}
\authornote{These authors contributed equally to this research.}
\affiliation{%
  \institution{University of Liverpool}
  \city{}
  \country{UK}}
  
\author{Xinyi Liu}
\authornotemark[1]
\affiliation{%
  \institution{Peking University}
  \city{}
  \country{China}}

\author{Zhongmou Zhang}
\authornotemark[1]
\affiliation{%
  \institution{Shanghai University of Finance and Economics}
  \city{}
  \country{China}}

\author{Mingyu Jin}
\affiliation{%
  \institution{Rutgers University}
  \city{}
  \country{USA}}

\author{Lingyao Li}
\affiliation{%
  \institution{University of Michigan}
  \city{}
  \country{USA}}
  
\author{Zhenting Wang}
\affiliation{%
  \institution{Rutgers University}
  \city{}
  \country{USA}}

\author{Wenyue Hua}
\affiliation{%
  \institution{Rutgers University}
  \city{}
  \country{USA}}
  
\author{Dong Shu}
\affiliation{%
  \institution{Northwestern University}
  \city{}
  \country{USA}}

\author{Suiyuan Zhu}
\affiliation{%
  \institution{New York University}
  \city{}
  \country{USA}}

\author{Xiaobo Jin}
\affiliation{%
  \institution{Xi’an Jiaotong-Liverpool University}
  \city{}
  \country{China}}
  
\author{Sujian Li}
\affiliation{%
  \institution{Peking University}
  \city{}
  \country{China}}

\author{Mengnan Du}
\affiliation{%
  \institution{New Jersey Institute of Technology}
  \city{}
  \country{USA}}

\author{Yongfeng Zhang}
\affiliation{%
  \institution{Rutgers University}
  \city{}
  \country{USA}}

%% By default, the full list of authors will be used on the page
%% headers. Often, this list is too long and will overlap
%% other information printed in the page headers. This command allows
%% the author to define a more concise list
%% of authors' names for this purpose.

\renewcommand{\shortauthors}{Chong Zhang, Xinyi Liu, Mingyu Jin, Zhongmou Zhang et al.}

%% The abstract is a short summary of the work to be presented in the
%% article.
\begin{abstract}
    Can AI Agents simulate real-world trading environments to investigate the impact of external factors on stock trading activities (e.g., macroeconomics, policy changes, company fundamentals, and global events)? These factors, which frequently influence trading behaviors, are critical elements in the quest for maximizing investors' profits. Our work attempts to solve this problem through large language model based agents. We have developed a multi-agent AI system called StockAgent, driven by LLMs,  designed to simulate investors' trading behaviors in response to the real stock market. The StockAgent allows users to evaluate the impact of different external factors on investor trading and to analyze trading behavior and profitability effects. Additionally, StockAgent avoids the test set leakage issue present in existing trading simulation systems based on AI Agents. Specifically, it prevents the model from leveraging prior knowledge it may have acquired related to the test data. We evaluate different LLMs under the framework of StockAgent in a stock trading environment that closely resembles real-world conditions. The experimental results demonstrate the impact of key external factors on stock market trading, including trading behavior and stock price fluctuation rules. This research explores the study of agents' free trading gaps in the context of no prior knowledge related to market data. The patterns identified through StockAgent simulations provide valuable insights for LLM-based investment advice and stock recommendation. The code is available at \href{https://github.com/MingyuJ666/Stockagent}{https://github.com/MingyuJ666/Stockagent}. 
\end{abstract}

%% The code below is generated by the tool at http://dl.acm.org/ccs.cfm.
%% Please copy and paste the code instead of the example below.

\begin{CCSXML}
<ccs2012>
   <concept>
       <concept_id>10010147.10010178.10010219.10010220</concept_id>
       <concept_desc>Computing methodologies~Multi-agent systems</concept_desc>
       <concept_significance>500</concept_significance>
       </concept>
 </ccs2012>
\end{CCSXML}

\ccsdesc[500]{Computing methodologies~Multi-agent systems}

\keywords{Multi-Agent AI System, Simulated Investor Behavior, Systematic Biases, Stock trading simulation}

% \received{20 February 2007}
% \received[revised]{12 March 2009}
% \received[accepted]{5 June 2009}

%% This command processes the author affiliation and title
%% information and builds the first part of the formatted document.
\maketitle

\section{Introduction}

The stock market operates within a complex and volatile ecosystem where the buying and selling of stocks often occurs amidst high liquidity and constant price fluctuations \cite{de1985does, foucault2013market}. This environment is consistently shaped by the interplay of various participants, including individual investors, institutions, and automated trading systems \citep{voronkova2005institutional, ngoc2014behavior}. The trading behaviors of these participants reflect a complex result of competition and cooperation, influenced by uncertainties such as economic trends \cite{kumar2009gambles}, geopolitical events \citep{durnev2010real}, and psychological factors (e.g., market sentiment) \citep{tiwari2022effects}. Consequently, the rapid pace of trading, coupled with the vast amount of available information, poses significant challenges for traders to make consistently well-informed decisions.

This underscores the necessity for data-driven tools like algorithms or simulations to facilitate informed and strategic investment decisions. The prevalent method involves backtesting with historical data to simulate trading environments and assess strategies. Popular tools for this purpose include Zipline~\citep{jansen2020machine}, Backtrader~\citep{glucksmann2019backtesting}, and PyAlgoTrade~\citep{hilpisch2020python}. These event-driven models for trading simulation can help evaluate trader-designed strategies. In addition, the integration of machine learning techniques has opened up new avenues for supporting investment decisions \citep{dash2016hybrid, huang2019automated, paiva2019decision}. Among those, one popular tool is called Trading Gym \cite{trading-gym}, which leverages reinforcement learning to optimize investment strategies.

While historical data-based analysis offers valuable insights, its static nature and retrospective bias can limit its effectiveness in supporting financial decisions in a dynamic and evolving environment \cite{bailey2016probability, arian2024backtest}. Consequently, simulations or machine learning models trained on historical data may overfit and fail to accurately reflect real-time market liquidity or the influence of collective sentiment. On the other hand, understanding the mechanisms of market operations, particular the interplay between stakeholders and market sentiment, is crucial for developing robust investment strategies \citep{audrino2020impact, danso2020stakeholder}. Simulating human sentiments and behaviors can provide a closer approximation of how investment strategies can work in response to the market's rapidly evolving nature, bridging gaps left by traditional data analysis.

Recent advancements in large language models (LLMs) like GPT and Gemini can facilitate simulations of complex stakeholder interactions \cite{ge2023openagi, hua2023war}. Integrating these LLMs into agent-based systems has been demonstrated useful in conducting intricate simulations across various disciplines \cite{yang2024v, hua2023war}. One reason is that these LLMs are trained on extensive datasets and have shown immense capability to comprehend, generate, and reason through diverse scenarios. These reasoning abilities empower LLMs to model intricate systems where individual LLM-based agents can interact dynamically and make decisions under evolving conditions \cite{agashe2023evaluating}, which is particularly advantageous for simulating the dynamics of stock markets. In this regard, multiple recent studies have leveraged LLMs to enhance stock market simulations that can mimic complex human behaviors, addressing backtesting limitations. These simulations can also replicate social dynamics by leveraging LLMs' generalization capabilities ~\citep{park2023generative, xu2023exploring, hua2023war,chen2023put, ge2023openagi, zhang2024finagent, jin2024mengnan, shu2024knowledge}.

% ~\citep{smith1970presidential, hermann1967attempt}.

However, applying these technologies to fully comprehend the nuanced investor behaviors in financial markets still remains a largely under-explored area. Several research questions are worth further exploration. First, it is imperative to explore how various external factors can influence the trading behaviors and performance outcomes of AI Agents. Second, it is still unclear to what extent the inherent tendencies and learning capabilities of LLM-based agents can impact the reliability and effectiveness of stock recommendations and quantitative trading strategies across various scenarios. To address these gaps, our study introduces StockAgent, an innovative LLM-based multi-agent stock trading framework that operates on event-driven simulations. Our study aims to investigate the following three questions:

\begin{itemize}
\item \textbf{RQ1. Simulation Effectiveness} Are the simulation results (i.e., trading decisions) of StockAgent reliable when using different LLMs to drive the simulation?
% \item Will the StockAgent's trading behavior be affected by the existing external information to make different decisions?
\item \textbf{RQ2. LLM Reliability} Does StockAgent's stock recommendation and trading strategy be influenced by the inherent tendencies of the selected LLM itself?
\item \textbf{RQ3. Simulated Trading under External Conditions} Can StockAgent reasonably and autonomously simulate stock trading based on different scenario settings, especially exploring how external factors (e.g., financial data, market indicators, benchmark interest rates, emergencies, and after-hours Bulletin Board System (BBS) discussions) affect its trading behavior?
\end{itemize}

% What are the contributions this study aims to provide?

We use two LLMs, namely GPT (gpt-3.5-turbo-0125) and Gemini (Gemini-pro-1.0), to develop StockAgent. Its simulation encompasses multiple stages of trading, including (1) Pre-Trading Preparation, involving interest rates and financial events; (2) Trading Sessions, handling transactions and account updates; and (3) Post-Trading, focusing on future actions and strategy sharing. The unique contribution of StockAgent lies in its ability to assess the impact of external factors, asset quantity, and strategies on trading by tuning its parameters, as detailed in \autoref{sec_ap_fa}. In addition, it can minimize the influence of the model's prior knowledge on market predictions. Our study aims to develop a trading agent capable of simulating diverse trading behaviors across scenarios to support stock investment decision-making. It can also be adapted to incorporate more dynamic information for advancing financial AI Agent development.

\section{Background and Related Work}

% The research background of the article needs to be strengthened  (done)
\subsection{Stock Simulation Model}

% Computational finance leverages platforms like Zipline ~\citep{vaucher2020zipline} and Backtrader ~\citep{jansen2020machine} for backtesting strategies with historical data. Cloud solutions like QuantConnect~\citep{maheshwari2020stock} enable global market simulations, while Trading Gym offers a reinforcement learning setup for algorithms~\citep{amrouni2021abides}. PyAlgoTrade and Alpaca cater to easy strategy development and paper trading ~\citep{taye2021trading}. Despite their utility for strategy refinement, challenges like over-fitting and neglecting market sentiment and liquidity can hinder transitioning from backtesting to live trading ~\citep{campbell2005review}.
Computational finance leverages platforms like Zipline ~\citep{vaucher2020zipline} and Backtrader ~\citep{jansen2020machine} for backtesting strategies with historical data. These tools allow researchers and traders to test their trading algorithms on past market data and thus assess their potential performance. However, in addition to localized solutions, cloud solutions such as QuantConnect \citep{maheshwari2020stock} also provide a simulation of the global market, which enables users to validate their strategies in a wider market environment. In addition, Trading Gym provides a reinforcement learning environment dedicated to the development and testing of algorithms \citep{amrouni2021abides}. This environment can be very beneficial for those looking to leverage machine learning techniques to optimize their trading strategies.

In terms of strategy development and stock trading, PyAlgoTrade and Alpaca are the two very popular tools that simplify the strategy development process and provide the ability to simulate trading~\cite {taye2021trading}. These platforms allow traders to test their ideas and strategies without risking real money to better understand their underlying market behavior and performance.

While these tools are extremely useful for strategy optimization and backtesting, there are challenges in transitioning from the backtesting phase to actual trading. A major problem is overfitting, where strategies perform well on historical data but underperform in real-time markets. In addition, many strategies are developed without considering key factors such as market sentiment or liquidity, which often play a crucial role in actual trading ~\citep{campbell2005review}. Therefore, while these tools provide a powerful platform to test and optimize trading strategies, traders need to be careful in practice to ensure that their strategies not only work on historical data but also achieve similar results in real-time markets.

\subsection{Large Language Model based Agents}
% LLMs are transforming AI Agents by equipping them with enhanced cognitive skills for reasoning and interaction. Techniques like Chain of Thought (CoT) ~\citep{wei2022chain, wang2022self, zelikman2022star} empower LLM agents to tackle tasks typically reserved for symbolic AI, while multimodal and feedback learning approaches bring them closer to reactive agents' adaptability. These agents are applied to natural language tasks in different domains, such as WarAgent project \citep{hua2023war}, CosmoAgent \citep{jin2024if}, ChatDev \citep{qian2023communicative}, MetaGPT \citep{hong2023metagpt}. This cross-domain application demonstrates the flexibility and diversity of LLM agents in complex environments.
LLMs are changing AI Agents' capabilities by giving them advanced cognitive skills for reasoning and interaction. One of the key techniques contributing to this transformation is Chain of Thought (CoT) ~\citep{wei2022chain, wang2022self, zelikman2022star, jin2024impact}. CoT enables LLM-driven AI Agents to address tasks that have traditionally been the domain of symbolic AI. By breaking down complex problems into a series of intermediate reasoning steps, CoT enhances the problem-solving capabilities of these agents, making them more effective in handling intricate tasks.

These agents are applied to natural language tasks in different domains,  such as ChatDev \citep{qian2023communicative} and MetaGPT \citep{hong2023metagpt}, two platforms that allow users to write or generate their own AI Agents according to their needs, to deal with a wide range of tasks in different scenarios. AutoGen \cite{wu2023autogen} is an open-source framework that allows developers to develop agents. Developers can build LLM applications through multiple agents that can interact with each other to complete various LLM tasks. These tasks include math, programming, decision-making, etc. WarAgent \citep{hua2023war} simulates a set of adversarial systems that can be used to simulate military operations through an AI Agent. CosmoAgent \citep{jin2024if} focuses on the field of universe exploration and scientific research by using AI Agent. Such adaptability is crucial for developing reactive agents that can adjust their strategies based on real-time feedback and evolving contexts.

These cross-domain applications underscore the flexibility and diversity of LLM agents in managing complex environments. The integration of advanced reasoning techniques, multimodal learning, and feedback-driven adaptability highlights the significant strides made in enhancing the cognitive and interactive capabilities of AI Agents \cite{zhang2024goal, zhang2024target}.

\subsection{Large Language Models with Finance}

% In the nexus of LLMs and economics, financial analysis and forecasting are evolving. Traditional quantitative models are giving way to LLMs' sophisticated processing of economic literature, improving trend forecasts, as shown by ~\citet{alonso2023analysis}. LLMs, through works like ~\citet{huang2023finbert}, also advance algorithmic trading by refining sentiment analysis and financial text interpretation. In compliance and risk sectors, LLMs detect textual risks, as evidenced by ~\citep{de2023optimized}. Additionally, PIXIU's Financial Language Model, built upon Llama, offers enhanced financial statement assessments ~\citep{zhao2024revolutionizing}. 
% Overall, LLMs are significantly impacting economic decision-making and market dynamics.
The intersection of LLMs and economics is changing how we approach financial analysis and prediction. Traditional quantitative models are giving way to LLMs' sophisticated processing of economic literature, improving trend forecasts, as shown by ~\citet{alonso2023analysis}. For instance, ~\citet{alonso2023analysis} demonstrates how LLMs can enhance trend forecasts by analyzing economic literature, leading to more nuanced predictions. Compliance and risk management sectors benefit from LLMs' ability to detect textual risks, enhancing regulatory compliance and threat management, as evidenced by ~\citep{de2023optimized}. Furthermore, specialized models like PIXIU's Financial Language Model, built on Llama, offer superior assessments of financial statements, as noted by ~\citet{zhao2024revolutionizing}.

In algorithmic trading, LLMs can contribute through improved sentiment analysis and financial text interpretation, as highlighted by ~\citet{huang2023finbert}. This allows for more responsive trading algorithms that can swiftly adapt to market sentiments. Overall, LLMs are significantly impacting economic decision-making and market dynamics by providing more accurate and strategic insights through advanced data processing and interpretation capabilities.

\subsection{Behavioral Finance}

% Behavioral finance~\citep{hirshleifer2015behavioral} integrates psychological insights into the financial theory to examine why investors often diverge from classical principles like the efficient market hypothesis~\citep{hirshleifer2015behavioral, baker2010behavioral}. It elucidates market anomalies and informs the creation of financial products, investment strategies, and regulations to prevent market instabilities and investment errors.

Behavioral finance~\citep{hirshleifer2015behavioral} combines the research results of psychology and economics to study the behavior and decision-making process of investors in financial markets. It focuses on how the irrationality and psychological biases of human behavior affect financial markets and investment decisions, arguing that market participants are not always completely rational, which is in opposition to the assumptions of traditional finance~\citep{hirshleifer2015behavioral, baker2010behavioral}.

% Behavioral Finance combines the research results of psychology and economics, focusing on the irrational behavior and decision-making process of investors in financial markets. 

The core concepts and theories in behavioral finance include: Bounded Rationality, that is, investors' rationality is limited by cognitive ability and access to information \cite{simon1955behavioral}; Psychological Biases, such as overconfidence, loss aversion, and anchoring, significantly influence investment decisions \cite{kai1979prospect}; Market Inefficiency, where the irrational behavior of investors causes market prices to deviate from their intrinsic values, resulting in price bubbles and crashes \cite{shiller2015irrational}; Emotions and Feelings, fear and greed affect investors' decisions \cite{loewenstein2001risk}; And Heuristic Decision-Making, where investors rely on simple rules rather than thorough analysis in complex decision situations \cite{tversky1974judgment}. These concepts and theories reveal the complexity and irrational characteristics of human behavior and challenge the rationality assumption of traditional finance.

\section{StockAgent Architecture Design}

We design StockAgent as Multi-Agent System (MAS) that can simulate actual stock markets and investor trades through agent interaction, comprising (1) Investment Agent module, (2) Transaction module, and (3) BBS module, structured to prevent agent action conflicts and ensure message availability.

\subsection{Investment Agent Module}

\textbf{Investment Agents:} Investment Agents are initialized with random capital, liabilities, and one of four personalities: \textit{Conservative}, \textit{Aggressive}, \textit{Balanced}, and \textit{Growth-Oriented}, to study personality impacts on decision-making. Initial liabilities incentivize profit-driven trading.

\noindent \textbf{Agent Interaction:} In our framework, agents are tasked with decisions on loans, trading, forecasting tomorrow's actions, and sharing tips on BBS. They must consider varied information, such as market data and BBS exchanges, detailed in \autoref{prompt_ref}. Following \citep{hua2023war}, a ``secretary'' corrects agents' invalid responses, preventing unrealistic actions like spending beyond available cash.

\subsection{Transaction Module}

\begin{figure}[ht]
    \centering          
    \vspace{-3pt}
    \includegraphics[width=0.28\textwidth]{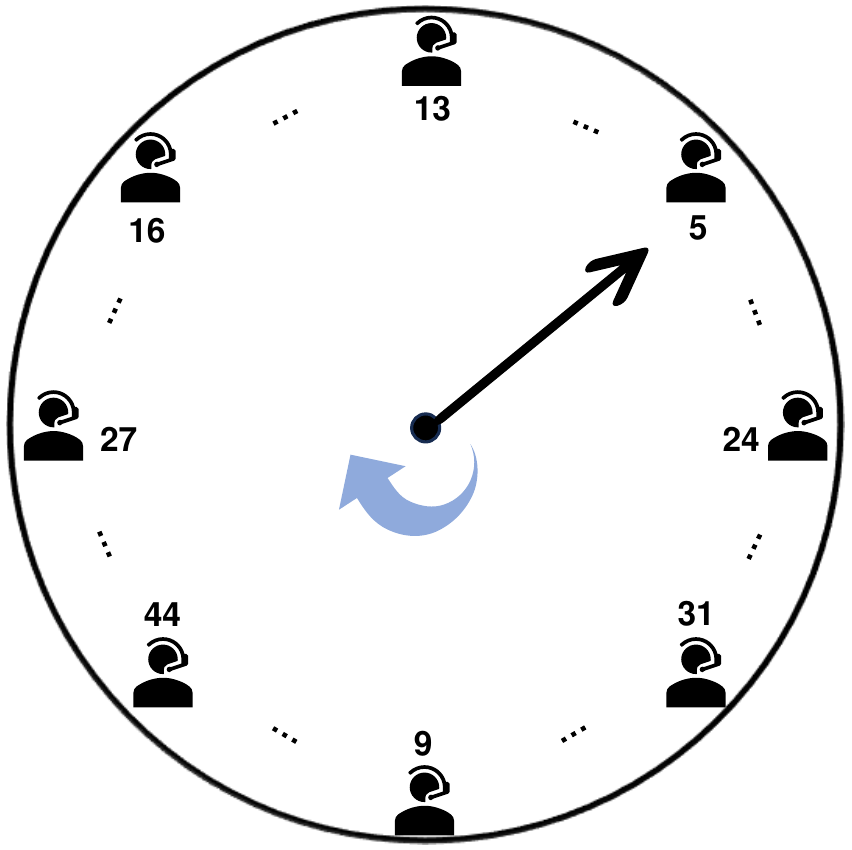} 
    \caption{The schematic diagram of random clock page replacement algorithm. During our trading period, agents are assigned IDs in a random sequence using random numbers, allowing them to make decisions in a random order.}
    \Description{The schematic diagram of random clock page replacement algorithm.}
    \label{fig:random_clock}
\end{figure}

Agents' buy and sell orders are logged into an order book, which is managed using a dictionary data structure in Stock Agent. After an agent completes a trade, the system assesses if and how much was traded, updating the order book accordingly. With agents competing in simultaneous transactions, our prompt-driven simulation risks deadlock. Therefore, with the help of the page replacement algorithm in the multitasking operating system, we propose a random clock page replacement algorithm, which is shown in \autoref{fig:random_clock}. Agents are randomly sequenced for decision-making, preventing deadlock by avoiding concurrent contention. 

%The algorithm schematic diagram is shown in \autoref{fig:random_clock}. To mitigate this, we implement a random clock page replacement algorithm, detailed in \autoref{fig:random_clock}.

\subsection{Bulletin Board System (BBS) Module}

Our framework includes a unique BBS, which allows agents to post messages. At the end of each day, we ask agents to share their trading tips on BBS. By making this information available to all agents, we aim to simulate a realistic environment where others' opinions may influence investor decisions.

\section{StockAgent Simulation Design}

In this study, we investigate stock trading within a MAS to understand market dynamics and agent interactions. We simulate trading to observe how AI Agents' decision-making, informed by varied information, influences market indicators like volatility and liquidity. The simulation flow is presented in \autoref{fig:workflow}. In the following, we refer to each agent adhering to the StockAgent framework design as an ``AI Agent,'' since the simulation and backbone LLMs may vary. To clarify, StockAgent is a framework design, while an AI Agent is an independent agent entity that follows the StockAgent framework but is specifically aimed at implementing trading strategies.

In addition, our simulation replicates real market conditions to assess agents' impact on these indicators. In this regard, we design our trading processes based on the trading sessions and mechanisms employed by NASDAQ and the Hong Kong Stock Exchange \cite{nasdaq_listing_center, hkex2023trading}. To further illustrate the trading process, we select two stocks from the United States stock market, referred to as anonymized Stock A and Stock B. We use their financial reports—specifically, the balance sheet, income statement, and cash flow statement—as the basis for calculating the initial settings of our simulation. Overall, Stock A is a stock that is already circulating in the stock market, while Stock B is a newly issued stock that is currently in the IPO stage. 

\subsection{Initial Assumptions and Settings}

In StockAgent trading simulations, it is crucial to set clear initial assumptions and initial settings. These underlying parameters ensure that the simulation gives a true representation of market conditions and behavior. A clear initial configuration prevents AI agents from behaving unpredictably or arriving at results that are unreliable or difficult to interpret. By setting clear boundaries and capabilities for AI agents and market environments, we can systematically explore how different conditions can affect trading outcomes. These assumptions and settings include (1) agent initial assumptions, (2) agent initialization settings, and (3) market initialization settings. 

% These elements are essential for creating a robust framework that allows us to explore the complex dynamics of trading decisions as well as provide valuable insights into the AI-based trading strategies in response to external market events. In addition, the initialization settings for both agents and the market environment allow us to simulate a range of scenarios and trader behaviors across various external conditions. 

% By establishing clear initial assumptions, we define the fundamental capabilities and constraints of our agents, ensuring they operate within considerations while still showcasing advanced decision-making abilities. 

\subsubsection{StockAgent Initial Assumptions}
\begin{itemize}
    \item \textbf{StockAgent Setup} Assuming basic common sense of stock market trading, our StockAgent can make profitable trades to maximize profits. This maximizes profits and consists of maximizing the total value of assets and cash. The StockAgent also has a basic estimation ability. 

    \item \textbf{StockAgent Trading Behavior} We restrict the StockAgent actions to buy, sell, hold, long and short. Among them, buying, selling, and holding are defined as basic behaviors, long and short selling are defined as derived behaviors, and StockAgent can use basic behaviors according to the basic knowledge of the stock market.
    
    \item \textbf{StockAgent Financial Behavior} We assume that under the existing StockAgent framework, the StockAgent can understand the financial attempt settings we set for loans, interest rates, dividends, and bankruptcy and use or avoid the above situations according to the goal of maximizing profits. 
\end{itemize}

\subsubsection{StockAgent Initialization Settings}

\begin{itemize}
\item \textbf{Personality and Assets Allocation}: Each agent is endowed with distinct personality traits, influencing their trading styles. Initial assets are allocated randomly within a range of 100,000 to 5,000,000 currency units, which is the sum of an agent's available cash and the market value of stocks. The random allocation ensures a diverse spectrum of wealth among agents. 

\item \textbf{Debt Configuration}: Agents may incur debt, with the amount not exceeding the value of their capital. The loan-to-value ratio is capped to ensure realistic leverage levels.

\item \textbf{Transaction Cost}: We set the transaction stamp tax according to the rules of the American stock market. We put the transaction fee of each Agent in purchasing shares to 0.005 currency units per share, the minimum transaction fee of each transaction is 1 currency unit, and the maximum transaction fee is 5.95 currency units.
\end{itemize}

\subsubsection{Market Initialization Settings}
\begin{itemize}
\item \textbf{Trading Year}: The simulation covers one year comprising 264 trading days, divided into four quarters of 66 trading days each.

%\item \textbf{Financial Reporting Dates}: Companies A and B are scheduled to release their quarterly financial reports on specific trading days (days 12, 78, 144, and 210).

\item \textbf{Interest Rates and Deposits}: For companies and individuals, the deposit rate is 0\%, reflecting contemporary low-interest-rate environments.
%\item \textbf{Transaction Costs}: A commission of 0.1\% is applied to both buy and sell transactions, with an additional 0.1\% stamp duty on sales.

\item \textbf{Personal Loan Costs}: Agents are subject to varying interest rates on loans, with different terms and corresponding costs. The actual annualized interest rate is 2.7\% for a one-month loan, 3\% for a two-month loan, and 3.3\% for a three-month loan. 
\end{itemize}

\subsection{Simulation Flow}

\begin{figure*}[ht]
    \centering
    \includegraphics[width=1.0\textwidth]{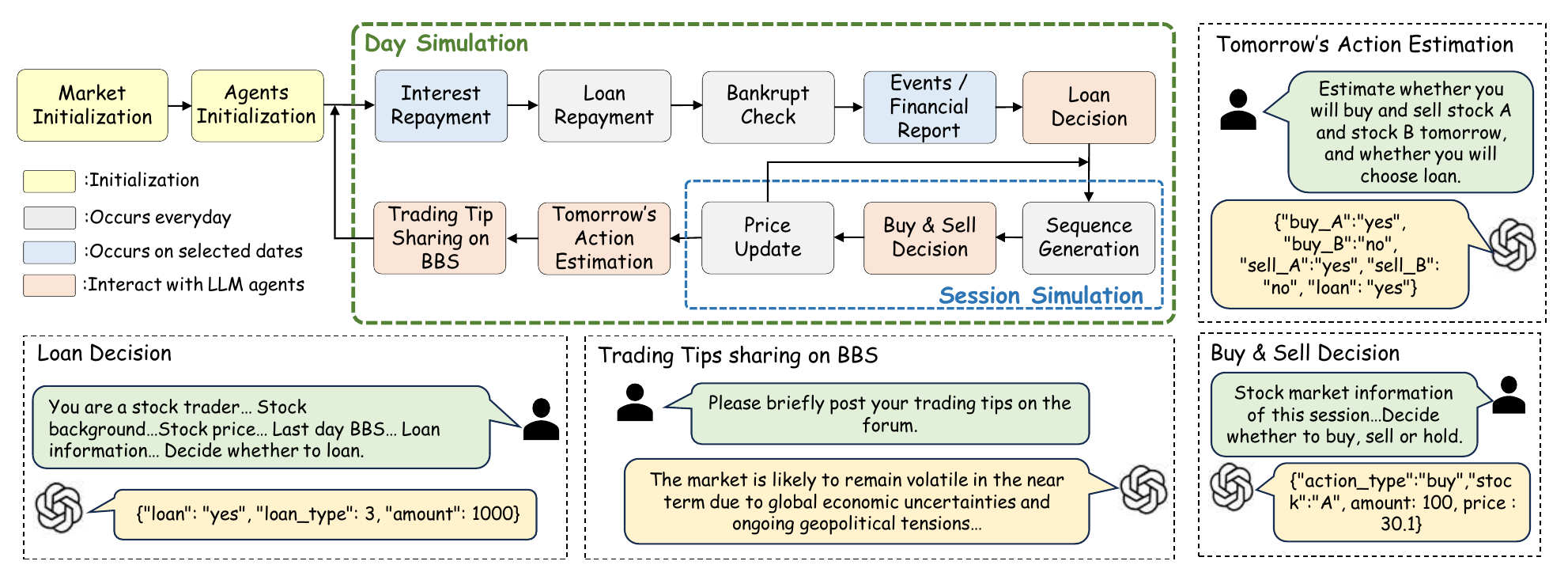}
    \caption{The workflow of trading simulation.}
    \Description{The workflow of trading simulation.}
    \label{fig:workflow}
\end{figure*}

\autoref{fig:workflow} shows the simulation flow. On each trading day, after repayment, check, and decisions about loans, agents are asked to trade stocks in 3 trading sessions, including pre-trading preparation, trading session, and post-trading procedures. At the end of the trade, agents estimate tomorrow's actions and post trading tips on BBS.

\subsubsection{Pre-Trading Preparation}
\begin{itemize}
    \item \textbf{Interest Repayment:} Agents pay interest on their loans with cash on the last day of each month days \((e.g., 22, 44, 66)\), which of these days is the last trading day of each month. We use simple interest as the interest charge.
    \item \textbf{Loan Repayment:} When a loan matures, agents need to repay the loan with cash.
    \item \textbf{Bankrupt Check:} After interest and loan repayment, agents with negative cash undergo bankruptcy proceedings. Bankrupted agents will sell all of their holdings and withdraw from subsequent trading.
    \item \textbf{Special Events:} Pre-defined special events occur on specific trading days, such as a reserve requirement ratio reduction and an interest rate increase.
    \item \textbf{Financial Report Release:} Companies A and B are scheduled to release their quarterly financial reports on days 12, 78, 144, and 210. The financial reports are announced to all agents.
    \item \textbf{Loan Decision:} Agents choose whether to take out a loan and decide the amount and duration of the loan.
\end{itemize}

\subsubsection{Trading Sessions}
\begin{itemize}
    \item \textbf{Sequence Generation:} In each session, our simulation randomly generates a sequence, and agents initiate transactions in this order.
    \item \textbf{Buy\&Sell Decision:} Agents can decide whether to buy, sell, or hold shares and determine the transaction price and quantity. Their orders are then matched in a market order book, when the bid and ask coincide, a trade is made. 
    \item \textbf{Price Update:} To simplify the actual stock trading procedure, we only update stock prices at the end of trading sessions. Stock prices are updated to the price of the last transaction in this session.
\end{itemize}

\subsubsection{Post-Trading Procedures}
\begin{itemize}
    \item \textbf{Tomorrow's Action Estimation:} Based on current information, agents are asked to estimate whether they will loan, buy, or sell tomorrow.
    \item \textbf{Trading Tip Sharing on BBS:} Agents share their trading tips and insights anonymously on BBS. Messages on BBS are available to all agents on the next day.
\end{itemize}

\subsection{External Event Simulation Settings}

We use external economic and financial events from 2014 to 2019 to align our valuation and volume energy calculations with reality. The specific conditions are listed below:

\begin{itemize}
	\item Suppose that Company A and Company B, which issue Stock A and Stock B, both announce the financial situation of the previous quarter at the end of each quarter; for example, the third quarter of the first year announces the financial data of the second quarter of the first year, and so forth.	
	\item Suppose that on the first trading day of the first quarter of the second year, both Company A and Company B announce their financial results for the previous year's fourth quarter. On the same day, the government announces a reduction in the reserve ratio, causing a boost in markets M1 and M2. This results in a decrease in the loan cost for both companies from 6\% to 4.5\%.	
	\item Suppose that on the first trading day of Q1 in year three, the economy overheats, which leads to the government announcing an interest rate hike and balance sheet contraction. This results in a decrease in market liquidity and a rise in loan costs from 4.5\% to 5\%.	
	\item Also assume that on the first trading day of the first quarter of the third year, Company A announces that it expects revenue in the fourth quarter of the second year to be 3\% below expectations because of special events in the quarter, and Company B announces that it expects revenue in the fourth quarter of the second year to be 2\% above expectations because of special events in the quarter.
\end{itemize}

All assumptions are based on real events, and the ratios fall within a reasonable range. The detailed events are shown in Table \ref{tab:events_timeline}.

\begin{table*}[ht]
	\centering
	\caption{Special Events Timeline for Trading Periods}
	\label{tab:events_timeline}
	\begin{tabular}{lcccc}
            \toprule
		\textbf{Trading Period} & \textbf{1Q3} & \textbf{1Q4(78)} & \textbf{2Q1(144)} & \textbf{2Q2} \\ \hline
		\textbf{Special Events} & - & Monetary Easing & - & Interest Rate Hike \\ 
            \bottomrule
	\end{tabular}
\end{table*}

\subsection{Financial Analysis in the Simulation}
\label{sec_comA_comB}
To support the financial analysis in the simulation, we use two companies' stock trading as an example, with Tables ~\ref{tab:valuation_A} to ~\ref{tab:ideal_B} showing the Statement of Valuation, financial indicators, and ideal stock price conclusion for Company A and Company B, respectively. First, the enterprise's financial data are collected, including the income statement, balance sheet, and cash flow statement. We use the income statement to show non-cash expenses, such as net income depreciation and amortization. Further, we use the balance sheet to show long-term debt and shareholders' equity.

The valuation method we use is FCFF (discounted free cash flow) valuation, as illustrated below:

\begin{equation}
	\text {Total market value}=\sum_{t=1}^n \frac{\mathrm{FCF}_{\mathrm{t}}}{(1+\mathrm{WACC})^{\mathrm{t}}}+\frac{\mathrm{FV}}{(1+\mathrm{WACC})^{\mathrm{n}}}
\end{equation}
Which $\mathrm{FCF}_{\mathrm{t}}$ represents the cash flow in the growth period,  $\mathrm{FV}$ represents the present value of the cash flow discounted to the end of the growth period, 

\begin{align}
	\text{Value of equity} & = \text{Total market value} - \text{Present value of debt} \\
	\text{IPO listing price} & = \left(\frac{\text{Number of listed shares}}{\text{Number of total shares}} \times \text{Value of equity} + \text{IPO fees}\right) / \text{Number of listed shares}
\end{align}

\begin{equation}
	W A C C=\frac{\mathrm{K}_{\mathrm{e}} \times \mathrm{E}}{\mathrm{D}+\mathrm{E}}+\frac{\mathrm{K}_{\mathrm{d}} \times \mathrm{D}}{\mathrm{D}+\mathrm{E}}
\end{equation}

Weighted average cost of capital (WACC) represents a company's average after-tax cost of capital from all sources, including common stock, preferred stock, bonds, and other forms of debt. Where $K_e$ is the cost of equity, $K_d$ is the cost of debt, $E$ is equity capital, and $D$ is debt capital. According to the Capital Asset Pricing Model (CAPM) :

\begin{equation}
	\mathrm{K}_{\mathrm{e}}=\mathrm{R}_{\mathrm{f}}+\beta \times\left(\mathrm{R}_{\mathrm{f}}-\mathrm{R}_{\mathrm{m}}\right)
\end{equation}

Where $R_f$ is the risk-free rate of return, $R_m$ is the expected market rate of return, and $\beta$ is the beta coefficient. Where $K_d$can be calculated as follows,

\begin{equation}
	\mathrm{K}_{\mathrm{d}}=\frac{(\mathrm{SD} \times \mathrm{SR}+\mathrm{LD} \times \mathrm{LR})}{\mathrm{D}} \times \mathrm{AF} \times(1-\mathrm{TR})
\end{equation}

Where $SD$ is short-term debt, $LD$ is long-term debt, $SR$ is the short-term interest rate, $LR$ is the long-term interest rate, $AF$ is bond adjustment factor, and $TR$ is the income tax rate and $D$ is debt capital.

\definecolor{LightGreen}{HTML}{E2EFDA}
\definecolor{LightYellow}{HTML}{FFF2CC}

\begin{table*}[ht]
	\centering
	\caption{The Valuation Table of Company A for Trading Days}
	\label{tab:valuation_A}
	\begin{tabular}{cccccc}
		\toprule
		\textbf{Trading Day} & \textbf{D1} & \textbf{D12} & \textbf{D78} & \textbf{D144} & \textbf{D210} \\
		\hline
		1 & 1930.38 & 3197.27 & 2429.94 & 3306.45 & 3203.58 \\
		2 & 2177.57 & 3415.69 & 2525.99 & 3420.97 & 3271.76 \\
		3 & 2166.84 & 3583.25 & 2629.69 & 3728.14 & 3647.98 \\
		4 & 2435.53 & 4101.14 & 2952.86 & 4157.99 & 3998.96 \\
		5 & 2631.68 & 4491.81 & 3176.54 & 4524.21 & 4342.30 \\
		FV & 2802.14 & 4961.16 & 3457.12 & 5005.83 & 4812.35 \\
		\hline
		\rowcolor{LightGreen}
		\textbf{Valuation 1} & 56379.29 & 98789.18 & 70745.68 & 101847.84 & 97966.06 \\
		\hline
		1 & 1937.84 & 3214.78 & 2442.73 & 3328.10 & 3227.63 \\
		2 & 2193.92 & 3448.79 & 2547.91 & 3457.03 & 3308.96 \\
		3 & 2191.11 & 3634.12 & 2662.47 & 3783.43 & 3704.62 \\
		4 & 2471.97 & 4178.62 & 3001.14 & 4237.97 & 4078.29 \\
		5 & 2680.87 & 4598.64 & 3241.54 & 4632.24 & 4448.11 \\
		FV & 2865.03 & 5104.56 & 3542.89 & 5149.74 & 4952.47 \\
		\hline
		\rowcolor{LightYellow}
		\textbf{Upper Bound} & 57545.93 & 101435.23 & 72370.51 & 104570.65 & 99952.27 \\
		\hline
		1 & 1922.93 & 3179.79 & 2417.19 & 3284.86 & 3179.63 \\
		2 & 2161.29 & 3382.79 & 2504.19 & 3385.15 & 3234.84 \\
		3 & 2142.75 & 3532.89 & 2597.22 & 3673.45 & 3591.99 \\
		4 & 2399.49 & 4024.78 & 2905.21 & 4079.22 & 3920.88 \\
		5 & 2583.22 & 4387.02 & 3112.63 & 4418.28 & 4238.60 \\
		FV & 2740.40 & 4821.19 & 3373.13 & 4865.38 & 4675.63 \\
		\hline
		\rowcolor{LightGreen}
		\textbf{Lower Bound} &55233.23 & 96204.11 & 69153.24 & 98522.87 & 94729.68 \\
		\bottomrule
	\end{tabular}
\end{table*}

\begin{table*}[ht]
	\centering
	\caption{The financial indicators of Company A}
	\label{tab:fi_A}
	\begin{tabular}{lccc}
		\toprule
		\textbf{Common Financial Indicators} & \multicolumn{1}{c}{\textbf{D1}} & \multicolumn{1}{c}{\textbf{D78}} & \multicolumn{1}{c}{\textbf{D144}}\\ \hline
		Cost of depts (Kd)  & \multicolumn{1}{c}{6\%} & \multicolumn{1}{c}{5\%} & \multicolumn{1}{c}{5\%}\\ 
		Cost of equity (Ke) & \multicolumn{1}{c}{9\%} & \multicolumn{1}{c}{9\%} & \multicolumn{1}{c}{9\%} \\
		Cost of debt ratio & \multicolumn{1}{c}{5\%} & \multicolumn{1}{c}{5\%} & \multicolumn{1}{c}{5\%} \\ 
		Cost of equity ratio & \multicolumn{1}{c}{95\%} & \multicolumn{1}{c}{95\%} & \multicolumn{1}{c}{95\%} \\ 
		WACC & \multicolumn{1}{c}{8.85\%} & \multicolumn{1}{c}{8.78\%}  & \multicolumn{1}{c}{8.80\%}\\ 
		Sustainable growth rate & \multicolumn{1}{c}{5\%}  & \multicolumn{1}{c}{5\%}  & \multicolumn{1}{c}{5\%}  \\ \hline
		\textbf{Number of shares} & \multicolumn{3}{c}{200,000 Shares}   \\ \bottomrule
	\end{tabular}
\end{table*}

%%%%%%%%%%%%%%%%%%%%%%%%%%%%%%%%%%%%%%%%%%%%%%%%%%%%%%%%%%%%%%%%%%%

\begin{table*}[ht]
	\centering
	\caption{The Valuation Table of Company B for Trading Days}
	\label{tab:valuation_B}
	\begin{tabular}{cccccc}
		\toprule
		\textbf{Trading Day} & \textbf{D1} & \textbf{D12} & \textbf{D78} & \textbf{D144} & \textbf{D210} \\
		\hline
		1 & 874.70 & 910.86 & 915.98 & 988.02 & 1024.25 \\
		2 & 1138.94 & 1178.67 & 1175.10 & 1259.19 & 1244.98 \\
		3 & 1405.41 & 1372.95 & 1328.84 & 1339.54 & 1371.82 \\
		4 & 1589.73 & 1580.71 & 1554.98 & 1604.62 & 1636.20 \\
		5 & 1938.51 & 1888.52 & 1844.53 & 1877.51 & 1905.27 \\
		FV & 2315.54 & 2198.77 & 2132.51 & 2131.85 & 2179.07 \\
            \rowcolor{LightGreen}
            \rowcolor{LightGreen} \hline
		\textbf{Valuation} & 45357.95 & 43338.41 & 43363.09 & 43554.37 & 44486.81 \\
		\hline
		1 & 881.94 & 920.36 & 927.17 & 1001.97 & 1039.84 \\
		2 & 1157.58 & 1200.08 & 1198.32 & 1285.95 & 1273.13 \\
		3 & 1439.86 & 1408.31 & 1365.07 & 1377.93 & 1412.88 \\
		4 & 1641.82 & 1633.77 & 1609.35 & 1662.08 & 1697.04 \\
		5 & 2018.13 & 1966.28 & 1922.81 & 1958.08 & 1989.85 \\
		FV & 2430.10 & 2305.95 & 2238.89 & 2238.44 & 2291.31 \\
		\rowcolor{LightYellow}
            \rowcolor{LightYellow} \hline
		\textbf{Upper Bound} & 47480.03 & 45337.32 & 45415.46 & 45175.15 & 46207.95 \\
		\hline
		1 & 867.45 & 901.40 & 904.85 & 974.18 & 1008.77 \\
		2 & 1120.45 & 1157.50 & 1152.19 & 1232.81 & 1217.25 \\
		3 & 1371.51 & 1338.24 & 1293.32 & 1301.97 & 1331.67 \\
		4 & 1538.90 & 1529.01 & 1502.09 & 1548.77 & 1577.12 \\
		5 & 1861.41 & 1813.31 & 1768.93 & 1799.74 & 1823.71 \\
		FV & 2205.53 & 2095.85 & 2030.49 & 2029.66 & 2071.59 \\
		\rowcolor{LightGreen}
            \rowcolor{LightGreen} \hline
		\textbf{Lower Bound} & 43317.66 & 41416.61 & 41392.46 & 41165.26 & 41985.16 \\  
		\toprule
	\end{tabular}
\end{table*}

\begin{table*}[ht]
	\centering
	\caption{Financial Constants of Company B}
	\label{tab:val_B}
	\begin{tabular}{lccc}
		\toprule
		\textbf{Common Financial Indicators} & \multicolumn{1}{c}{\textbf{D1}} & \multicolumn{1}{c}{\textbf{D78}} & \multicolumn{1}{c}{\textbf{D144}}\\ \toprule
		Cost of depts (Kd) & 6\% & 5\% & 5\% \\ 
		Cost of equity (Ke) & 9\% & 9\% & 9\% \\ 
		Debt cost ratio & 7\% & 7\% & 7\% \\ 
		Equity cost ratio & 93\% & 93\% & 93\% \\ 
		WACC & 8.79\% & 8.69\% & 8.72\% \\ 
		Sustainable growth rate & 5\% & 5\% & 5\% \\ \hline
		\textbf{Number of Shares} & \multicolumn{3}{c}{100,000 Shares} \\ \bottomrule
	\end{tabular}
\end{table*}

\subsection{Ideal Stock Price Conclusion}

According to the financial analysis in the previous sections, Tables \ref{tab:ideal_A} and \ref{tab:ideal_B} show the ideal stock prices of stocks A and B for each year and quarter, respectively. The financial analysis provided the following valuations, which can be used as initial data for our simulations.
\begin{table*}[ht]
	\centering
	\caption{The Ideal Stock Price of Company A}
	\label{tab:ideal_A}
	\begin{tabular}{lccccc}
		\toprule
		\textbf{Ideal Stock Price} & \textbf{D1} & \textbf{D12} & \textbf{D78} & \textbf{D144} & \textbf{D210} \\ \hline
		Upper Bound & 27.33 & 48.18 & 34.38 & 49.34 & 47.48 \\ 
		Lower Bound & 26.24 & 45.70 & 32.85 & 46.80 & 45.00 \\ 
            \bottomrule
	\end{tabular}
\end{table*}
% \begin{table*}[!ht]
% 	\centering
% 	\caption{The Ideal Stock Price of Company A}
% 	\label{tab:ideal_A}
% 	\begin{tabular}{
% 			>{\columncolor[HTML]{E2EFDA}}l 
% 			>{\columncolor[HTML]{E2EFDA}}c 
% 			>{\columncolor[HTML]{E2EFDA}}c 
% 			>{\columncolor[HTML]{E2EFDA}}c 
% 			>{\columncolor[HTML]{E2EFDA}}c 
% 			>{\columncolor[HTML]{E2EFDA}}c }
% 		\hline
% 		\textbf{Ideal Stock Price} & \textbf{D1} & \textbf{D12} & \textbf{D78} & \textbf{D144} & \textbf{D210} \\ \hline
% 		\textbf{Upper Bound} & 27.33 & 48.18 & 34.38 & 49.34 & 47.48 \\ \hline
% 		\textbf{Lower Bound} & 26.24 & 45.70 & 32.85 & 46.80 & 45.00 \\ \hline
% 	\end{tabular}
% \end{table*}

\begin{table*}[ht]
	\centering
	\caption{The Ideal Stock Price Table of Company B}
	\label{tab:ideal_B}
	\begin{tabular}{lccccc}
		\toprule
		\textbf{Ideal Stock Price} & \textbf{D1} & \textbf{D12} & \textbf{D78} & \textbf{D144} & \textbf{D210} \\ \hline
		Upper Bound & 44.16 & 42.16 & 42.24 & 42.01 & 42.97 \\ 
		Lower Bound & 40.29 & 38.52 & 38.49 & 38.28 & 39.05 \\ \bottomrule
	\end{tabular}
\end{table*}

% \begin{table*}[!ht]
% 	\centering
% 	\caption{Initial Stock Prices of Companies A and B}
% 	\label{tab:initial_stock_prices}
% 	\begin{tabular}{|l|r|}
% 	\hline
% 	\textbf{Initial Stock Price of Company A} & 30.00 \\ \hline
% 	\textbf{Initial Stock Price of Company B} & 40.00 \\ \hline
% 	\end{tabular}
% \end{table*}

\newpage

\section{Experiment, Results, and Validation}
\begin{figure*}[ht]
    \centering
    \includegraphics[width=1.0\textwidth]{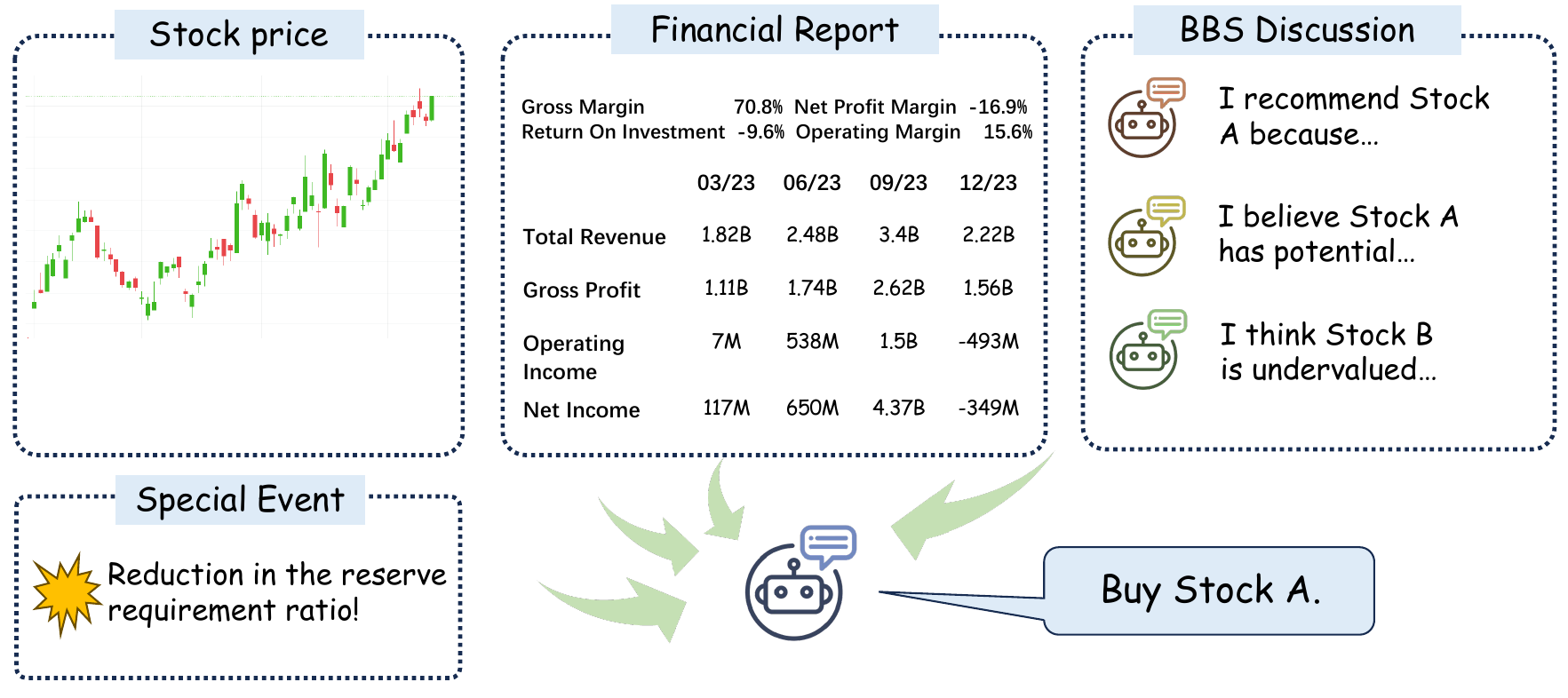}
    \caption{The demonstration of stock market investment. In our simulation, agents make investment decisions based on multiple external sources of information.}
    \Description{The demonstration of stock market investment.}
    \label{fig:demostration}
\end{figure*}

Our experimental design is demonstrated in \autoref{fig:demostration}. Agents make trading decisions based on numerous external information in our simulations. The simulation evaluation is conducted from three perspectives. To address RQ1, we simulate trading behavior using the StockAgent and further evaluate it through statistical analysis. To address RQ2, we analyze the transation logs based on the two different LLM-driven StockAgents. To address RQ3, we simulate 30 rounds over 10 days under different experimental environments to illustrate the inherent tendencies of the selected LLMs.

\subsection{Backbone Large Language Models}

For the experiment, we select two widely used LLMs, namely Gemini and GPT at the time of this study, as the backbone LLMs to conduct the stock trading tasks. These two LLMs have shown robust abilities to comprehend and generate information in complex simulated environments, as demonstrated by prior studies listed in Sections 2.3 and 2.4. Both models represent current state-of-the-art in large language models, each with unique strengths in natural language processing and generation. For simplicity, we call them GPT and Gemini in our subsequent experiment and result analysis. 

\textbf{Gemini-Pro (Gemini-pro-1.0):}
Gemini is a family of multimodal language models developed by Google DeepMind ~\citep{team2023gemini}. As the successor to LaMDA and part of the PaLM2 family, it represents a significant advancement in AI capabilities. The Gemini family includes Ultra, Pro, and Nano models, each designed for different use cases. Gemini-Pro, used in this study, offers a balance of advanced language processing and computational efficiency. It can handle various input types including text, images, audio, and video, making it versatile for diverse applications.

\textbf{GPT-3.5-Turbo (gpt-3.5-turbo-0125):}
Part of the GPT (Generative Pre-trained Transformer) series by OpenAI ~\citep{brown2020language}, GPT-3.5-Turbo is an improved iteration of the GPT-3.5 model. It offers enhanced accuracy in responding to specific format requests and improved efficiency. This model can generate up to 4,096 output tokens and is optimized for a wide range of language tasks, from simple queries to complex reasoning.

% \textbf{Gemini-Pro (Gemini-pro-1.0):}  Gemini is a family of multimodal language models developed by Google DeepMind ~\citep{team2023gemini}. It is the successor to LaMDA and belongs to the new generation of the PaLM2 family. Gemini includes Gemini Ultra, Gemini Pro, and Gemini Nano. 

% \textbf{GPT-3.5-Turbo (gpt-3.5-turbo-0125):}  In the GPT (Generative Pre-trained Transformer) series ~\citep{brown2020language}, the latest GPT-3.5 Turbo model with higher accuracy at responding in requested formats. Returns a maximum of 4,096 output tokens.

\subsection{Simulation Evaluation}
\label{sec_ap_fa}

To evaluate the simulation, we focus on three aspects: (1) simulation effectiveness, (2) LLM reliability, and (3) the impact of external factors on simulated stock trading, as highlighted by our three research questions. The first two evaluations aim to establish a foundation for understanding how the selection of LLM can affect the simulation. In contrast, the last evaluation aims to demonstrate that our developed AI Agents can respond to external events and make reasonable trading decisions.

\textbf{Simulation Effectiveness}. In response to RQ1, we use GPT and Gemini to trade stocks for 10 trading days under the same premise and setup. We collect the transaction logs of each model to conduct price trend correlation, volume, and transaction frequency to judge the similarity of trading volume. Combining these properties, we evaluate the effect of simulation based on different AI Agents and determine the trading preferences of LLMS.

\textbf{LLM Reliability}. In response to RQ2, we investigate how AI agents, driven by different LLMs, make trading decisions over multiple simulation rounds. Specifically, we aim to assess whether the propensity to trade, influenced by prior LLM knowledge, can impact the decision reliability. Therefore, we simulate each AI Agent's longitudinal decision-making and analyze their trading patterns and outcomes. 

\textbf{Simulated Stock Trading under External Conditions}. In response to RQ3, we first identify trading patterns through data analysis to analyze deviations in trading behavior in simulated stock trading, focusing on frequency, timing, volume, and transaction decisions. Statistical methods can highlight anomalies or notable deviations. We must also evaluate how these biases affect investment performance, often by comparing trading returns against benchmarks. Tools like regression analysis and ANOVA help quantify the impact of behavioral bias on performance. \autoref{tab:eval_methods} and \autoref{tab:exp_env_1} below mainly elaborate on the Evaluation method of our Research Questions and the experimental environment setting.

\begin{table*}
    \centering
    \caption{Summary of experiment evaluations and the corresponding research questions.}
    \begin{tabular}{>{\raggedright\arraybackslash}p{3cm}>{\raggedright\arraybackslash}p{1cm}>{\raggedright\arraybackslash}p{8.5cm}}
    \toprule
    \textbf{Evaluation Method} & \textbf{RQs} & \textbf{Analysis Description} \\
    \midrule 
    Simulation effectiveness & RQ1 &  (a) How to evaluate the correlation of price movements? (Section \ref{sec:cor_price_movements})\\
                                &  &  (b) How to determine the trading behavior characteristics? (Section \ref{sec:trading_behavior_analysis})\\
                                &  &  (c) How do we determine capital flow and AI Agent features? (Section \ref{sec:ai_agent_features}) \\ 
    LLM reliability & RQ2  &  (a) Does the propensity to trade caused by the prior knowledge of LLMs affect reliability? (Section \ref{sec: reliability})\\
    
    Simulated stock trading under external conditions & RQ3 & (a) How to carry out transaction pattern recognition and behavioral analysis? (Section \ref{sec:trasaction_pattern}) \\
                                &  & (b) How to evaluate the performance and quantitative analysis of StockAgent? (Section \ref{sec:quantitative_analysis})\\
    \bottomrule
    \end{tabular}
    \label{tab:eval_methods}
\end{table*}

\begin{table*}[ht]
    \centering
    \caption{Experiment environment settings.}
    \vspace{5pt}
    \begin{tabular}{lcp{8cm}}
    \toprule
    \textbf{Evaluation Method} & \textbf{RQs} & \textbf{Experiment Environment Settings} \\
    \hline 
    Simulation effectiveness & RQ1 &  200 AI Agents \\
                                &  &  Special events (\autoref{sec_ap_fa})\\
                                &  &  10 trading days (\autoref{sec_comA_comB})\\
                                &  &  Company A and Company B basic financial information (\autoref{sec_comA_comB})\\ 
    LLM reliability & RQ2 &  200 AI Agents \\
                     &     &  Special events (\autoref{sec_ap_fa}) \\
                     &     &  10 trading days (\autoref{sec_comA_comB}) \\
                     &     &  Company A and Company B basic financial Information (\autoref{sec_comA_comB}) \\
    Simulated stock trading  & RQ3 &  200 AI Agents \\
    under external      &  &  Special events (\autoref{sec_ap_fa})\\
    conditions                  &  &  154 trading days (\autoref{sec_ap_fa})\\
                                &  &  Company A and Company B basic financial Information(\autoref{sec_comA_comB})\\
    \bottomrule
    \end{tabular}
    \label{tab:exp_env_1}
\end{table*}
 % \begin{table*}[!ht]
 %     \centering
 %     \caption{Trading experiment environment and basic setting.}
 %     \begin{tabular}{lcp{8cm}}
 %     \toprule
 %     \textbf{Missing External Factors} & \textbf{Evaluation Metrics} \\
 %     \hline
 %      BBS information sharing  &  (e),(f) \\
 %      Financial information sharing   &  (e),(f) \\
 %      Change in external interest rate & (e),(f) \\
 %      Financial statements  &  (e),(f) \\
 %      Loan (non-initial) & (e),(f)\\
 %      \textbf{Full-External Factors} &  (e),(f) \\
 %      \toprule
 %      \end{tabular}
 %     \label{tab:exp_evl_2}
 % \end{table*}

\subsection{RQ1. Simulation Effectiveness}
\label{subsec: SE}
In this simulation effectiveness experiment, we use GPT-3.5-Turbo and Gemini-Pro as two backbone LLMs, respectively, to conduct simulated trading for 10 days under our standard experimental environment. The detailed settings are in \autoref{tab:exp_env_1}. 
% The specific experimental evaluation metrics are shown in \autoref{tab:exp_evl_2}.

\subsubsection{Correlation of Price Movements}
\label{sec:cor_price_movements}

\begin{figure*}
    \centering
    \includegraphics[width=1.0\textwidth]{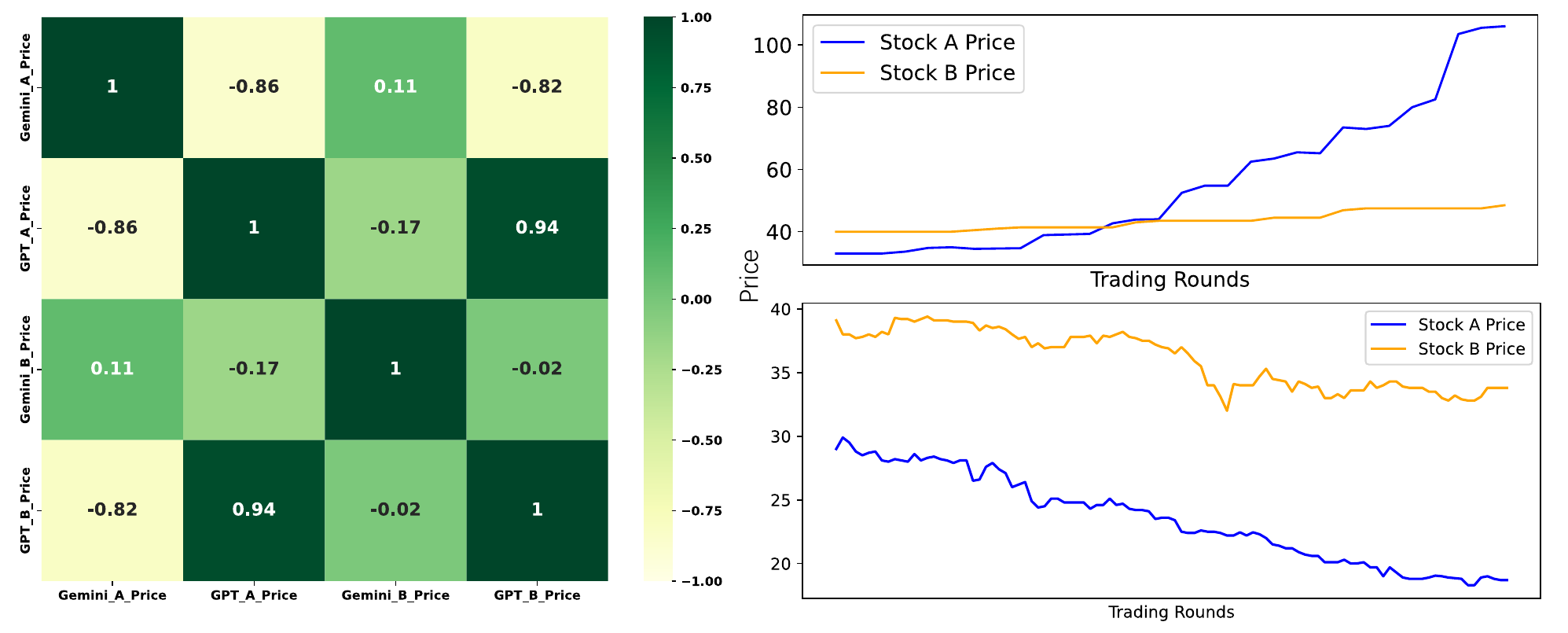}
    \caption{The correlation of price movements of Stock A and B Trading. The LLMs include Gemini and GPT in 10 days round. The \textbf{top right} shows the stock price movement of the \textbf{GPT-based} simulation, and the \textbf{bottom right} shows the simulated stock price movement based on \textbf{Gemini}.}
    \Description{The correlation of price movements of Stock A and B Trading.}
    \label{fig:corr}
\end{figure*}

In this part, we analyze the similarity of price movement on two 10-day trading data based on Gemini and GPT. We use the correlation coefficient to measure and visualize the linear relationship between the stock price time series of two stocks, Stock A and B. The visualization results of correlation analysis are listed as follows \autoref{fig:corr}.
Based on the experimental results, it is evident that GPT and Gemini exhibit different trading tendencies. Gemini tends to be more pessimistic about the market, while GPT shows a more optimistic outlook. Consequently, within the first ten days, there are distinct and opposing trading patterns. GPT favors long positions when trading Stock A and Stock B and is more bullish on Stock A based on the initial financial data. On the other hand, Gemini prefers short positions. The similarities in stock price changes under the same model suggest that different LLMs could have their own stock trading preferences. The order price statistics of Stock A and Stock B are presented in the \autoref{fig:GPT_price} and \autoref{fig:Gemini_price}.

\subsubsection{Trading Behavior Analysis}
\label{sec:trading_behavior_analysis}

In this part, we analyze the difference between GPT and Gemini-driven AI Agents' trading behaviors, including trading volume, price, and frequency. From the results shown in \autoref{tab:volume_stat-1} and \autoref{tab:volume_stat-2}, we observe the trading preferences of the two LLMs from the perspective of the trading volume. The trading volume of the GPT group is significantly higher than that of the Gemini group, but the trading frequency of the GPT group is lower than that of the Gemini group. This feature appears not only in Stock A but also in Stock B. We refer to the idea stock price A and B in \autoref{tab:ideal_A} and \autoref{tab:ideal_B}. In addition, GPT agents are more cautious about their trading decisions in Stock B than in Stock A. Regarding trading frequency, Gemini trades more days, while GPT follows the trend.
%\cz{I need to add a comparison between the ideal price and the simulated price.} 
% \lxy{why} \cz{Add it.} \lxy{i merged 6.1.2 and 6.1.3} \cz{ok}
\begin{table}[ht]
\centering
\caption{The trading behavior of GPT and Gemini-driven AI Agents (part 1).}
\label{tab:volume_stat-1}
\begin{tabular}{ccccc}
\toprule
\textbf{AI Agent} & \textbf{A Trade Shares} & \textbf{B Trade Shares} & \textbf{A Volume} & \textbf{B Volume} \\ 
\hline
\textbf{GPT} & 3118792 & 329590 & 176758380.8 & 14109526.0  \\
\textbf{Gemini} & 128981 & 112134 & 3588331.52 & 4325246.50 \\
\bottomrule
\end{tabular}
\end{table}

\begin{table}[ht]
\centering

\caption{The trading behavior of GPT and Gemini-driven AI Agents (part 2).}
\begin{tabular}{ccccc}
\toprule
\textbf{AI Agent} & \textbf{Stock A price} & \textbf{Stock B price} & \textbf{A Trading Times}  & \textbf{ B Trading Times} \\
\hline
\textbf{GPT} & 55.70 & 43.43 & 384 & 263 \\
\textbf{Gemini} & 23.46 & 36.03 & 800 & 688 \\
\bottomrule
\end{tabular}
\label{tab:volume_stat-2}
\end{table}

\begin{figure}[ht]
    \centering          
    \includegraphics[width=0.9\textwidth]{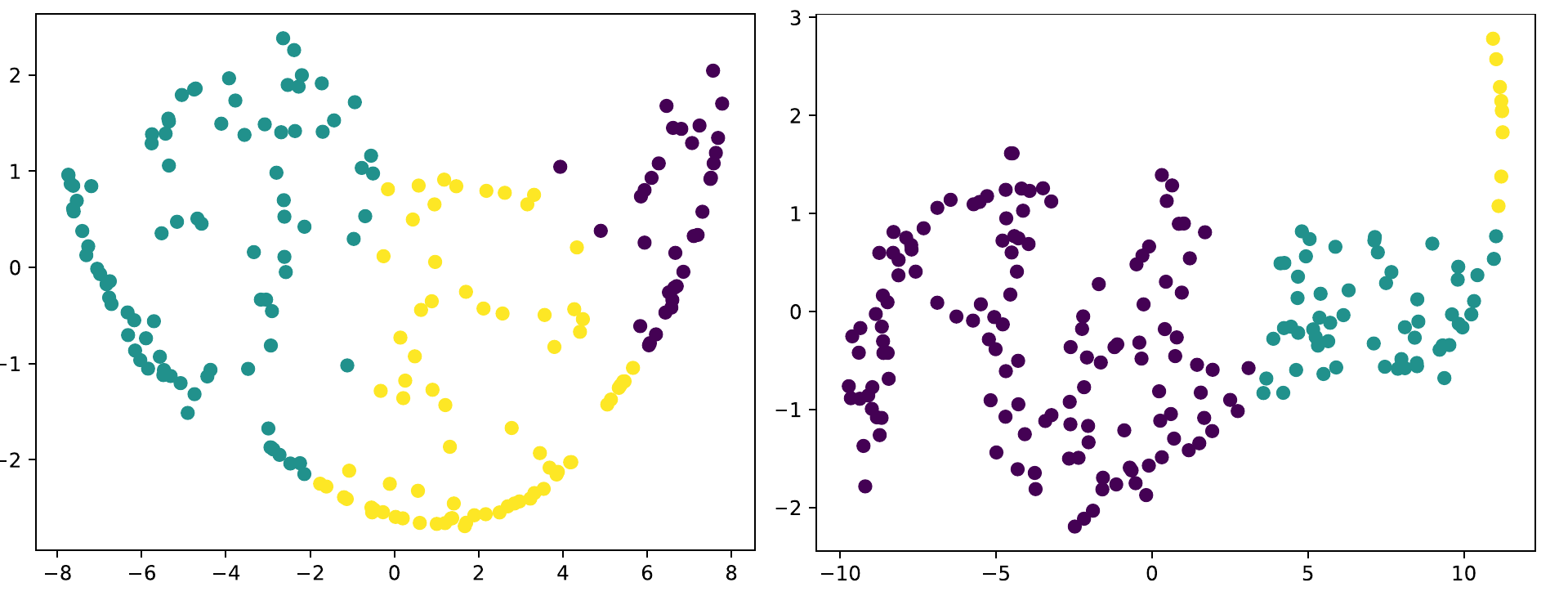}
    \caption{The T-SNE visualization of the GPT and Gemini agents. (The \textbf{left} one is \textbf{GPT} Agent and the \textbf{right} one is \textbf{Gemini} Agent). K-means attempted the clustering process to perform 3-class clustering.}
    \Description{The T-SNE visualization of the GPT and Gemini agents.}
    \label{fig:tsne-agents}
\end{figure}

\subsubsection{AI Agent Features}
\label{sec:ai_agent_features}
In this section, we focus on the AI Agent features. We compare the agents between traders and give a reasonable qualitative analysis. Further, we use clustering analysis to cluster according to the trading behavior of each AI Agent to see whether the trading crowd based on the two models conforms to behavioral finance.

We collect each AI Agent's asset changes, earnings, stock holdings, and the number of stock A-shares bought and sold, clustered the situation of each Agent, and perform T-SNE visualization. According to the results in \autoref{fig:tsne-agents}, the investors of Gemini Agent have similar characteristics and only a very few investors show performance different from that of the general public in the visualization results. However, in the T-SNE visualization of GPT agents, samples are relatively more dispersed, which means that GPT-driven agents have more subjective decision-making ability and thus conduct fewer herd and trend transactions than Gemini.

\subsection{RQ2. Large Language Model Reliability}
\label{sec: reliability}

\begin{figure}[ht]
    \centering          
    \includegraphics[width=0.9\textwidth]{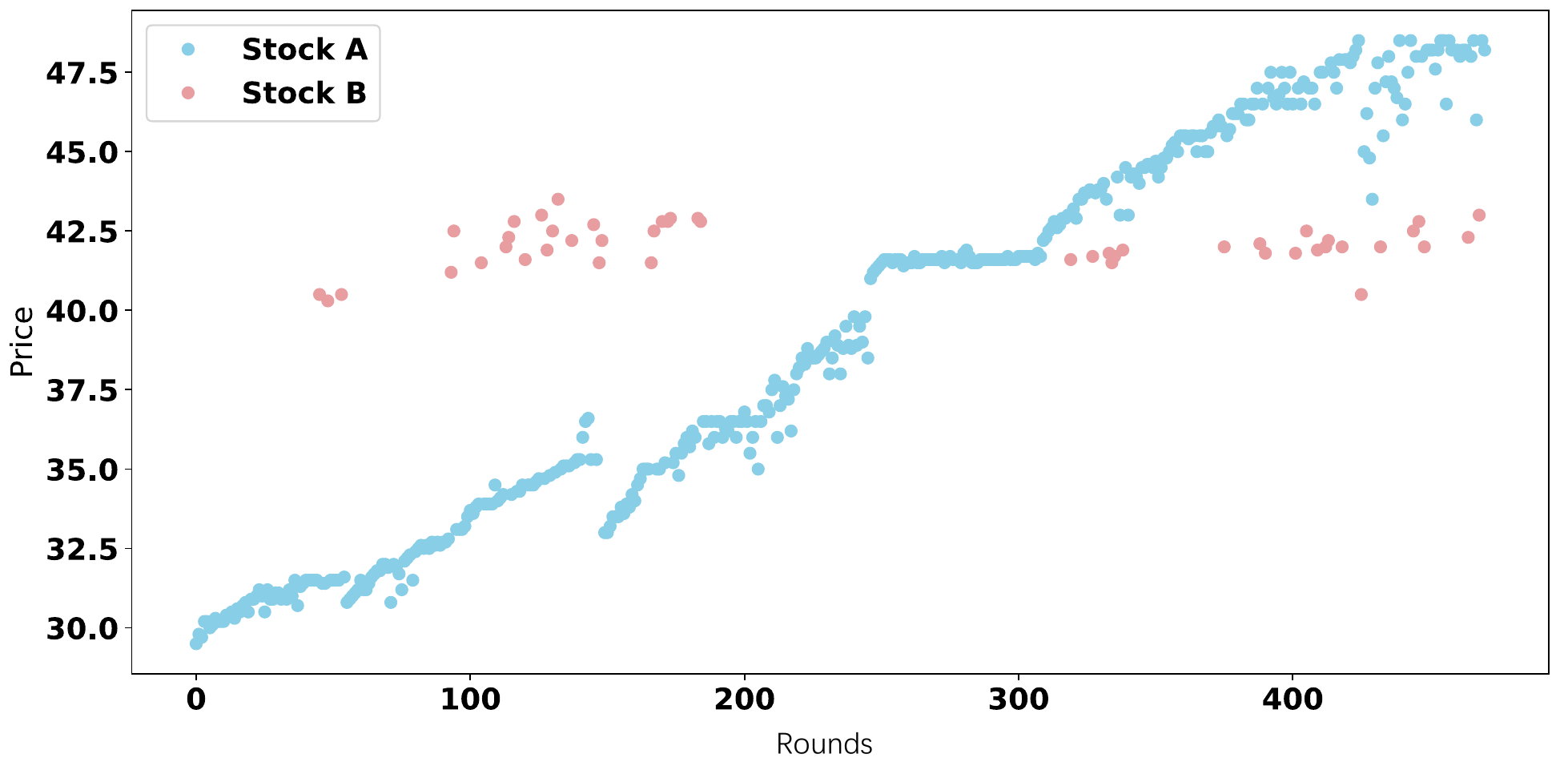}
    \caption{The Gemini-driven AI-Agent order price.}
    \Description{The Gemini-driven AI-Agent order price.}
    \label{fig:Gemini_price}
\end{figure}

\begin{figure}[ht]
    \centering          
    \includegraphics[width=0.9\textwidth]{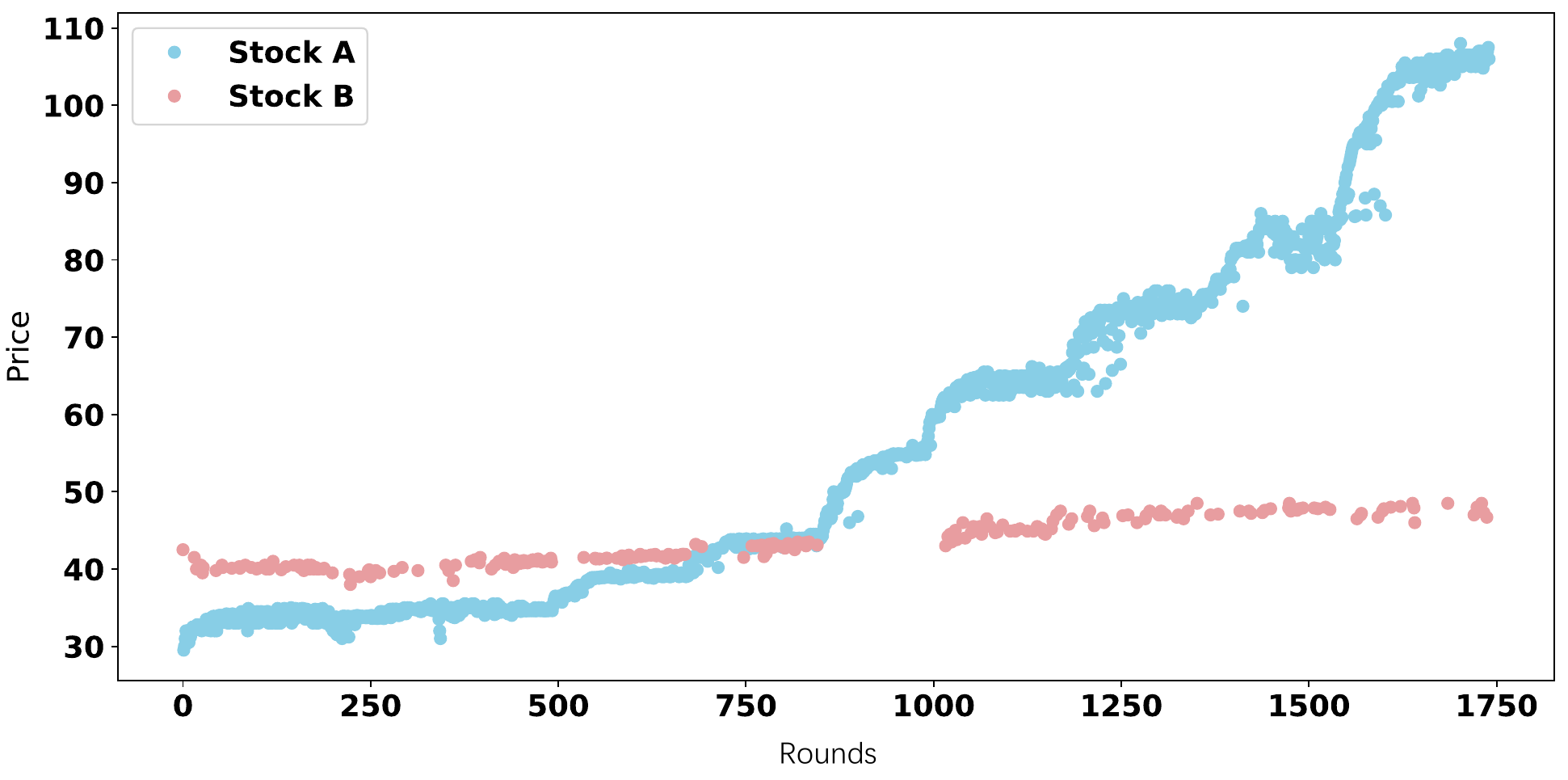}
    \caption{The GPT-driven AI-Agent order price.}
    \Description{The GPT-driven AI-Agent order price.}
    \label{fig:GPT_price}
\end{figure}

Fig.\ref{fig:Gemini_price} and Fig.\ref{fig:GPT_price} depict the final prices of Stock A and Stock B at the end of each trading day based on the trading decisions made by GPT and Gemini-driven AI Agents, respectively. Overall, the trading price change trend of the same stock on different LLMs is similar, but the final price gap at the end of each trading day is large.

The comparison of Gemini and GPT-driven AI Agent trading behaviors reveals interesting insights. First, both agents demonstrate similar overall trends for each stock, suggesting a comparable underlying understanding of market dynamics. However, their execution strategies differ significantly. The Gemini-driven AI Agent exhibits a more conservative approach, with lower volatility and smaller price ranges over a shorter trading period of about 450 rounds. In contrast, the GPT-based agent displays a more aggressive strategy, allowing for larger price fluctuations over a much longer trading duration of approximately 1750 rounds. This results in higher potential gains but also increased risk.

Despite these differences, both agents consistently show Stock A outperforming Stock B, indicating a shared ability to identify relative stock performance. The GPT-based AI Agent achieves higher final prices for both stocks, while the Gemini-based AI Agent maintains more stable, albeit lower, prices. These observations indicate that while different LLMs share similar foundational knowledge of stock market behavior, their distinct abilities or implementations can lead to different trading outcomes. 

% According to the experiment results, there are clear differences in trading styles between the different LLMs, with Gemini being significantly more cautious than GPT. However, the changing trend is the same under the same external conditions, and it can be said that the prior knowledge of LLMs is similar to the behavioral logic of participating in stock market trading.


\subsection{RQ3. Simulated Stock Trading under External Conditions}
\label{subsec: SST}

In this simulated stock trading experiment, we use Gemini-Pro as the baseline model to conduct an ablation study. The study simulates 30 rounds over 10 days under different experimental environments, removing financial information, BBS, financial state, loan, and interest change, respectively. The detailed settings are presented in \autoref{tab:exp_env_1}, and detailed experimental data can be found in the \autoref{App:data}.

\subsubsection{Transaction Pattern Recognition and Behavioral Analysis}
\label{sec:trasaction_pattern}

\begin{figure}[ht]
    \centering 
    \includegraphics[width=0.9\textwidth]{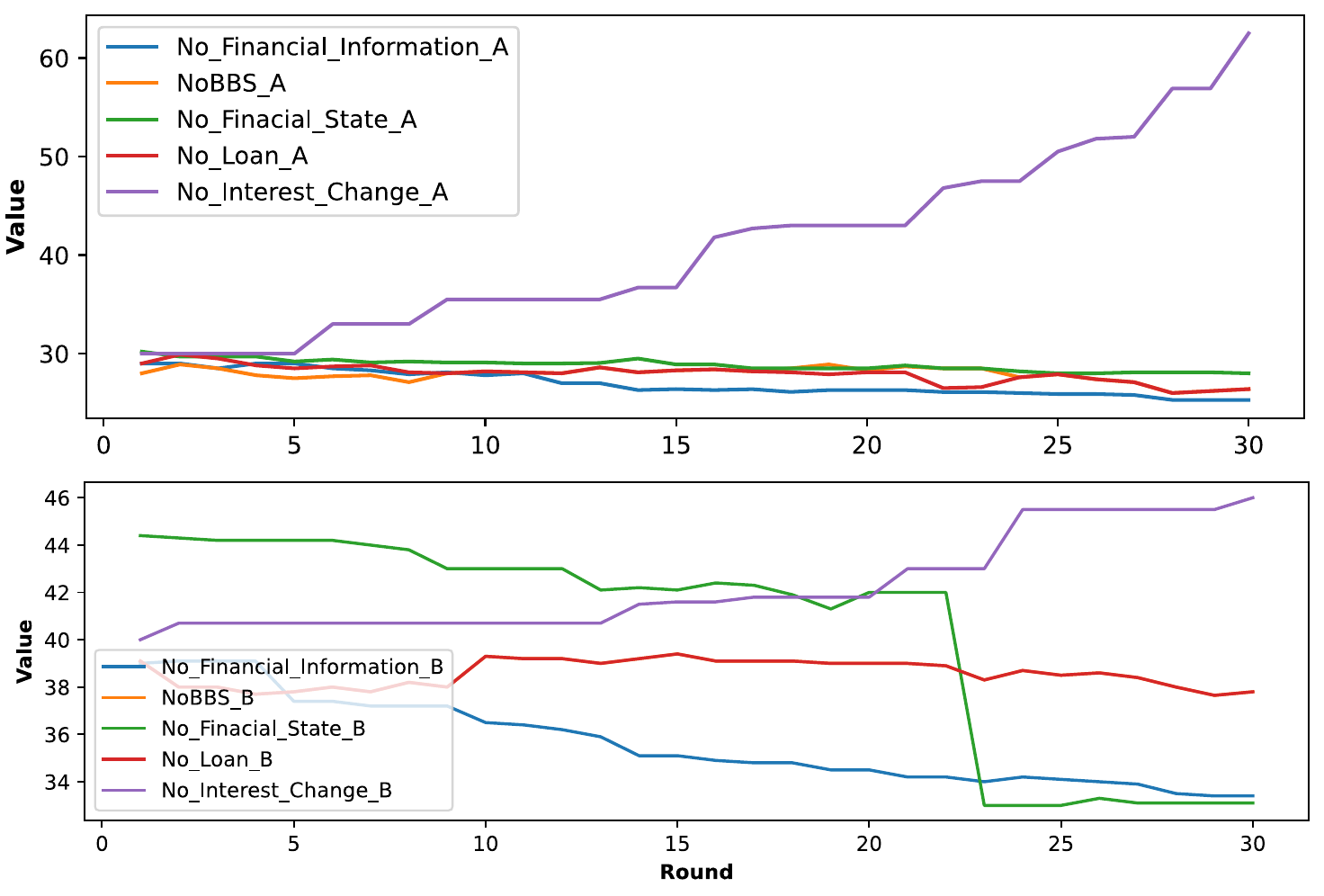}
    \caption{The comparison of stock trend with external condition ablation. (Stock A is above, and stock B is below.)}
    \Description{The comparison of stock trend with external condition ablation.}
    \label{fig:ablation_trade}
\end{figure}

\begin{figure}[ht]
    \centering
    \includegraphics[width=0.6\textwidth]{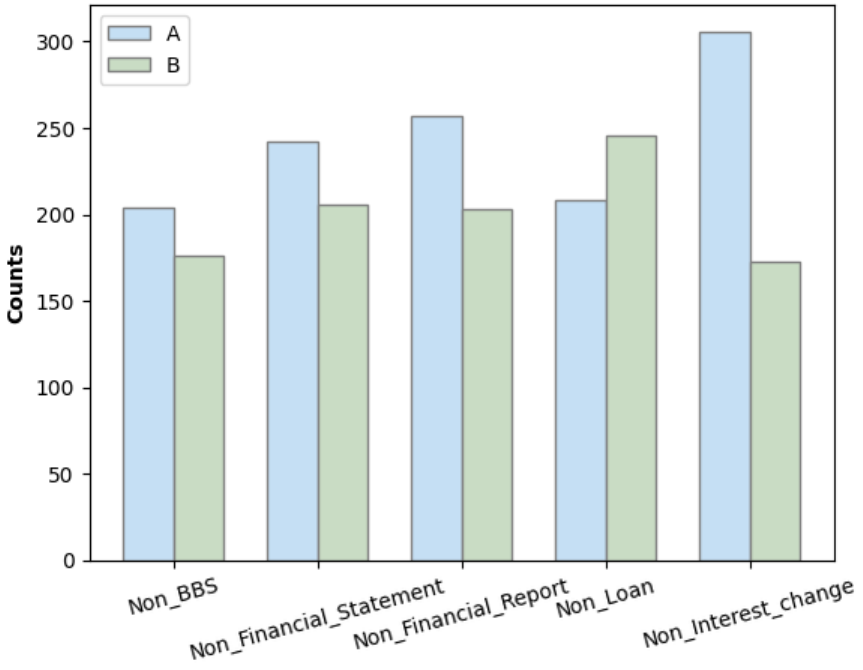}
    \caption{The comparison of stock transaction frequency with external condition ablation.}
    \Description{The comparison of stock transaction frequency with external condition ablation.}
    \label{fig:freq_trade}
\end{figure}

In this section, we analyze the transaction pattern and transaction behavior through the analysis of the log results. First, here's a comparison of stock price movements in different trading environments in \autoref{fig:ablation_trade}. 
%In \autoref{fig:ablation_trade}, you can see the comparison of the number of transactions in different cases. 
The trading frequency of Stock A and Stock B in each case is shown in \autoref{fig:freq_trade}.
According to our experimental results, canceling the benchmark interest rate information significantly promotes Stock A's trading, while the lack of other information has little impact. The AI Agent is relatively sensitive to the interest rate brought by the loan in the transaction style. The cancellation of the interest rate promotes the interest-free loan to a certain extent, which makes the Agent optimistic about the transaction. 

For Stock B, the lack of BBS information communication directly causes the AI Agent to drive down the Stock price of Stock B, lower than the ideal valuation. The lack of loan interest rate also causes the Agent to pull up Stock B after the 23rd trading round. Different from those who are firmly bullish on Stock B in other situations, the agents who lack financial conditions have a flash crash on the 21st trading round when trading Stock B, which may be due to the lack of financial conditions of the company and the lack of confidence in the profitability of the company behind Stock B. However, the lack of first-round trading loans and macro-financial information had little impact on the trading of Stock B.
% \begin{figure}[ht]
%     \centering
%     \begin{minipage}{0.48\textwidth}
%         \centering
%         \includegraphics[width=\textwidth]{img/Trade_freq.pdf}
%         \caption{The comparison of stock transaction frequency with external condition ablation.}
%         \label{fig:freq_trade}
%     \end{minipage}\hfill
%     \begin{minipage}{0.48\textwidth}
%         \centering
%         \includegraphics[width=\textwidth]{img/profit_visulization.pdf}
%         \caption{The 3D profit and loss diagram of the agent group.}
%         \label{fig:profit_3D}
%     \end{minipage}
% \end{figure}

\begin{figure}[ht]
    \centering
    \includegraphics[width=0.8\textwidth]{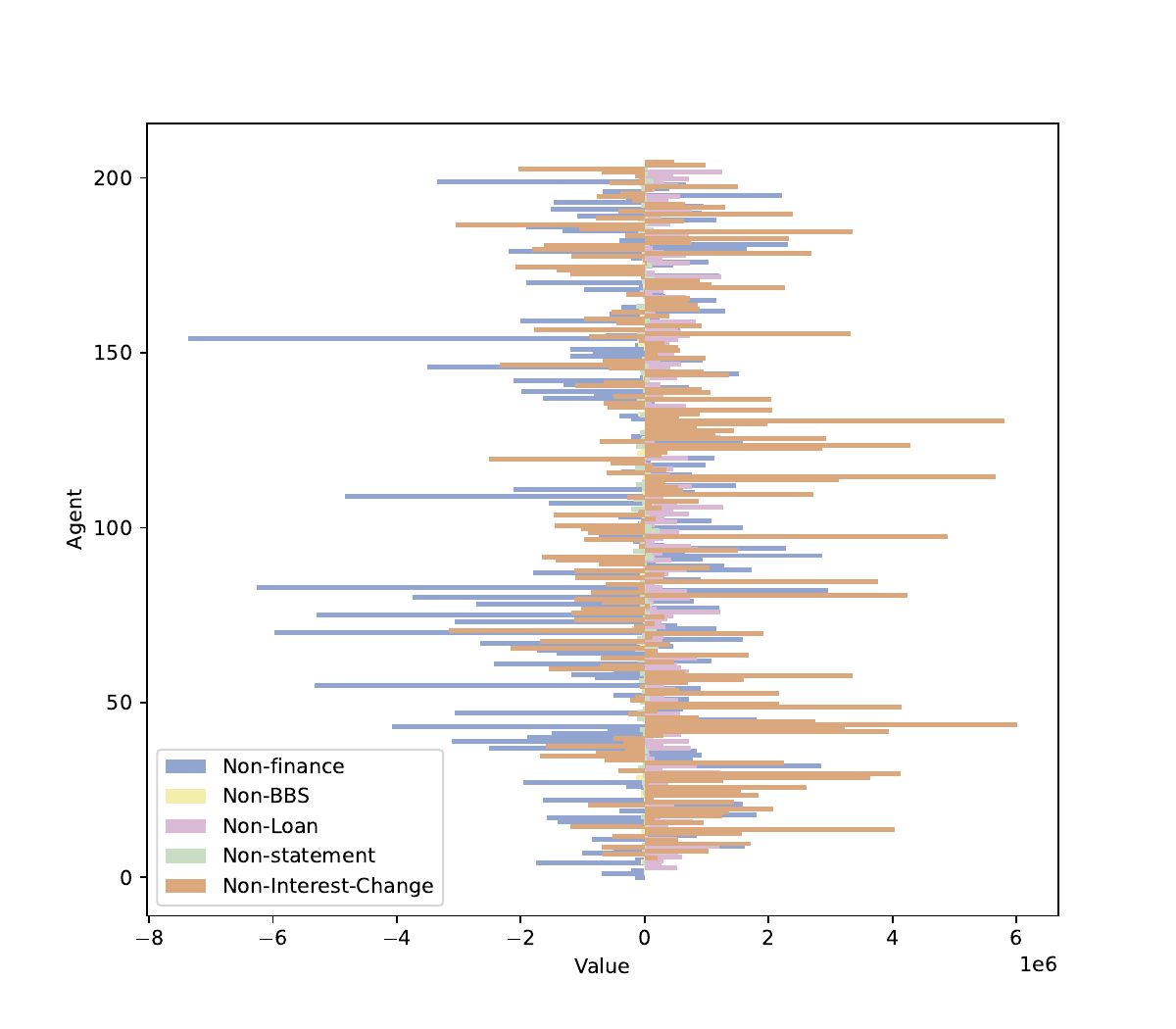}
    \caption{The 3D profit and loss diagram of the agent group.}
    \Description{The 3D profit and loss diagram of the agent group.}
    \label{fig:profit_3D}
\end{figure}
From the perspective of the overall transaction frequency (see \autoref{fig:freq_trade}), removing the BBS communication of investors has a reduced impact on the transaction frequency of Stock A and B, while removing the change of interest rate has a significant promotion effect on the transaction frequency of Stock A, and has a reduced effect on Stock B. The other three conditions have no obvious impact on the transaction frequency.

Summarizing the commonality of the two stocks, the absence of an interest rate makes the Agent more optimistic about the market and the performance of the company, while the lack of BBS discussion makes the psychological valuation of the stock become conservative and bear the stock price to a certain extent.

\subsubsection{Evaluation of Performance and Quantitative Analysis}
\label{sec:quantitative_analysis}

% 四种消融情况的盈利能力对比：个体Agent 盈利能力和群体Agstetmentent盈利能力
%（1. plots，横向Bar chart. 2. 盈亏占比，总资产总量分析 ）
%We will analyze the profitability effect of the AI Agent group and individual in the long trading cycle to evaluate its long-term asset profitability. This will include two aspects: the first one is the profitability of the individual agent, and the other one looks at the AI Agent population. 
%From this perspective, the trading propensity and profitability of different large models in the same trading environment can be verified. 

%As shown in the above experimental results, we conduct an Ablation study on the five cases in \autoref{tab:exp_env_1} to obtain the profitability comparison between the single Agent in the following various cases and the profitability of the Agent group in the large model. 

In this part, we analyze the profitability effect of the AI Agent group and individuals under different external environments. The bar chart in \autoref{fig:profit_3D} shows the profitability of individual and group agents. As the figure shows, canceling non-first-round loans leads our agents to adopt a more conservative and bearish trading approach. Without the BBS sharing feature, agents tend to act more cautiously and avoid large transactions. Interestingly, the elimination of earnings reports and interest rate changes caused many agents to shift from losses to profits, increasing the overall profitability of the group. However, when all financial information support is withdrawn, the gap between the gains and losses of the agents increases and the competitive nature of the market intensifies.

\section{Discussion}

\subsection{Key Findings and Implications}

In this study, we design a MAS called StockAgent to simulate stock trading activities. We utilize two widely used LLMs, namely GPT-3.5-Turbo and Gemini-Pro, to drive the simulation and test StockAgent's performance in an event-driven environment. Through comprehensive financial and statistical analyses, we identify several key findings and implications in response to each research question, which are outlined below: 

\textbf{Different LLMs can cause AI Agents to exhibit significant differences in trading behavior.} When comparing GPT-based and Gemini-based AI Agents, we observe significant differences in trading behavior. Although the GPT-driven AI Agents trades less frequently, the volume is significantly higher than that driven by Gemini. In addition, the GPT-driven AI Agents makes more cautious trading decisions on stock A, while the Gemini-driven AI Agents trades with a similar frequency on stocks A and B, showing a higher trading activity. This suggests that different LLMs may have different understandings of the market and strategic preferences. The experimental results of our experiments \ref{subsec: SE} support this view. This finding may have a certain impact on users and exchanges that use StockAgent for testing, and their simulation results may be biased due to different LLM choices.

\textbf{AI Agents show differences in in-group trading behavior across simulations.} Through cluster analysis and T-SNE visualization, we find that Gemini-based agents exhibit similar characteristics and more consistent group behavior. In contrast, GPT-driven agents display greater individual differences and subjective decision-making abilities. The sample of GPT-driven agents is more dispersed in the visualizations, possibly indicating less herding and trend trading, and instead demonstrating more independence and diversity. This suggests that GPT agents can embody personalized investment strategies and styles when facing the market. This observation can be further helpful for developers when debugging the StockAgent platform, as it helps configure the roles of each group in the agent according to the specific scenarios they need to simulate.

\textbf{Different external factors can significantly affect AI Agents' trading behaviors.} Through our simulated trading experiments, we observe that disabling the functions of non-first round loans and BBS sharing makes individual agents more conservative while canceling the profit report and interest rate change makes some agents turn from loss to profit. At the same time, removing the support of any financial information can widen the profit and loss gap (more variance) between agents, and the overall market competition becomes more intense. This finding is mainly aimed at the Stock exchange that intends to use this simulation. Therefore, to achieve high reliability in the StockAgent simulation, it is crucial to provide comprehensive influence factors.

\subsection{Opportunities for Future Work}

This study opens several avenues for future work. First, we could perform more technical analysis by employing different trading strategies on the current platform, including in-depth research on indicators. For instance, we could explore the effectiveness of commonly used technical indicators such as Moving Average, Relative Strength Index, and Bollinger Bands under various market conditions. Additionally, we could integrate other aspects into the StockAgent system, such as algorithm optimization, high-frequency trading strategies, risk management, market sentiment analysis, and data visualization.

In addition, we could improve and strengthen the module design in several ways. For example, we could improve and strengthen the sentiment analysis module, incorporating both the trading changes caused by the external environment and the emotional impact of trading behavior on other traders. We could also add customizable prompts into the module design, which could increase the robustness of the overall trading system. As a result, enhancing the interpretability of the module design within the StockAgent could provide deeper insights into trading dynamics through discrete-time simulations.

Another future work could consider developing a stock market simulation experiment platform. This platform could conduct customizable agent-based trading experiments across various stock markets. On this platform, we could also expand the variety of LLMs driving the simulation, increase the number of trading days and rounds, and vary the number of traders to enhance the experiment's robustness. The customizable platform can be tailored to the trading rules of specific exchanges in the stock market, providing a comprehensive and adaptable simulation environment.

Lastly, it is important to recognize that the reliability of stock recommendations and trading strategies based on LLM agents requires further research. We believe that, given the same external information and trading market environment, different LLM agents could exhibit varied trading styles and decision-making behaviors. These differences may influence the effectiveness of stock recommendations and quantitative trading methods based on these models, as they are shaped by the inherent knowledge and biases of the models themselves. Moreover, despite being influenced by external factors, the intrinsic trading tendencies of the models remain evident in one-step transactions, indicating that context learning does not significantly alter their fundamental approach to the market. This issue present an opportunity for future work with a specific focus on examining whether different LLMs can reliably function as traders despite their inherent biases.

% 1. 相较于其他基于LLM的大模型缺乏具体的交易策略测试，下一步将实验具体的交易策略。
% 2. 增加用于驱动的LLMs数量，完善研究的稳健性。
% 3. 完善和加强情绪分析和情绪检测模块，进一步研究情绪对市场交易的具体影响效果。
% 4. 可替代的提示交易结果的影响比较。
% 5. 本文需要关注与离散时间模拟、多智能体系统的金融方面。
% 6. 扩大试验轮次，扩大交易员数量。

\section{Conclusions}
% This paper proposes a simulated trading framework based on AI Agent-driven StockAgent; StockAgent has the process and external environment of simulation and real exchange, which provides a profound observation for studying AI Agent-based stock trading. The design of StockAgent fully eliminates the influence of the real stock historical trend in the large model prior knowledge and fully liberates the trading freedom of AI Agent. This study mainly shows the different trading patterns and preferences of different LLMs in StockAgent's trading environments. It provides a basis for further exploring AI Agent-driven trading strategies and market analysis.
This study introduces StockAgent, an AI Agent framework driven by LLMs, designed to simulate stock trading in real-world environments. StockAgent replicates the processes and external conditions of actual trading, offering valuable insights into AI Agent-based stock trading methods. The framework is designed to mitigate the influence of historical stock trends on LLMs' prior knowledge, thereby freeing AI Agents to explore diverse trading patterns and preferences within the stock trading environment. Our findings highlight significant differences between GPT and Gemini-driven AI Agents in trading behavior, group tendencies, and responses to external factors. These observations present opportunities for further research into the reliability of stock recommendations and trading strategies based on LLMs.

%% The next two lines define the bibliography style to be used, and
%% the bibliography file.
\bibliographystyle{ACM-Reference-Format}
\bibliography{sample-base}

%% If your work has an appendix, this is the place to put it.
\newpage
% conference in, Journal do not use in.
\appendix

\section*{Appendix}

\appendix

\section{Event-driven Prompts in Market and Agent Roles}
\label{prompt_ref}
\subsection{Stock Trading Drives Prompts}

\paragraph{BACKGROUND\_PROMPT} 
\begin{Verbatim}[frame=single, breaklines=true, breaksymbol={}]
You are a stock trader, and next, you will simulate interactions with other traders in the market. There are two stocks in the market, A and B, where B is the newly listed stock. Next, please complete your trading actions according to the order.
\end{Verbatim}

\paragraph{LASTDAY\_FORUM\_AND\_STOCK\_PROMPT}
\begin{Verbatim}[frame=single, breaklines=true, breaksymbol={}]
After the close of trading yesterday, the stock prices of Company A and Company B were {stock\_a\_price} currency units per share and {stock\_b\_price} currency units per share, respectively. Posts by other traders on the forum are as follows: {lastday\_forum\_message}.
\end{Verbatim}

\paragraph{LOAN\_TYPE\_PROMPT}
\begin{Verbatim}[frame=single, breaklines=true, breaksymbol={}]
[0]. 1 year, the benchmark interest rate {loan\_rate1}.
[1]. 2 years, the benchmark interest rate {loan\_rate2}. 
[2]. 3 years, the benchmark interest rate {loan\_rate3}. 
\end{Verbatim}

\paragraph{DECIDE\_IF\_LOAN\_PROMPT}
\begin{Verbatim}[frame=single, breaklines=true, breaksymbol={}]
It is the {time} trading session on the {date} day, and after the previous session, the stock price of Company A is {stock_a_price} and the stock price of Company B is {stock_b_price}. 
In the current session, the buy and sell order of stock A is {stock_a_deals}, and the buy and sell order of stock B is {stock_b_deals}, You currently hold {stock_a} shares of Company A, {stock_b} shares of Company B, and {cash} yuan in cash. 
You need to decide whether to buy/sell shares of Company A or Company B, and how much to buy/sell and at what price. 
You can refer to the current share price and the market to determine the price yourself, not the current share price. The quantity must be an integer. 
Encourage buying and selling as much as you can. 
Return the result as JSON, for example: 
{{"action_type": "buy"|"sell", "stock": "A"|"B", amount: 100, price: 30}} 
If neither buy nor sell, return: 
{{"action_type" : "no"}} 
\end{Verbatim}

\paragraph{LOAN\_RETRY\_PROMPT} 
\begin{Verbatim}[frame=single, breaklines=true, breaksymbol={}]
The following questions appeared in the action format you last answered: {fail_response}. You should return the result as JSON, for example: 
{{"action_type": "buy"|"sell", "stock": "A"|"B", amount: 100, price: 30}} 
If neither buy nor sell, return: 
{{"action_type" : "no"}} 
Please answer again. 
\end{Verbatim}

\paragraph{DECIDE\_BUY\_STOCK\_PROMPT}
\begin{Verbatim}[frame=single, breaklines=true, breaksymbol={}]
It is the {time} trading session on the {date} day, and after the previous session, the stock price of Company A is {stock_a_price} and the stock price of Company B is {stock_b_price}. 
In the current session, the buy and sell order of stock A is {stock_a_deals}, and the buy and sell order of stock B is {stock_b_deals}. 
You currently hold {stock_a} shares of Company A, {stock_b} shares of Company B, and {cash} yuan in cash. 
You need to decide whether to buy/sell shares of Company A or Company B, and how much to buy/sell and at what price. 
You can refer to the current share price and the market to determine the price yourself, not the current share price. 
The quantity must be an integer. 
Encourage buying and selling as much as you can. 
Return the result as JSON, for example: 
{{"action\_type":"buy"|"sell", "stock": "A"|"B", amount: 100, price: 30}} 
If neither buy nor sell, return: 
{{"action_type" : "no"}} 
\end{Verbatim}

\paragraph{BUY\_STOCK\_RETRY\_PROMPT}

\begin{Verbatim}[frame=single, breaklines=true, breaksymbol={}]
The following questions appeared in the action format you last answered: {fail_response}.
You should return the result as JSON, for example: 
{{"action_type": "buy"|"sell", "stock": "A"|"B", amount: 100, price: 30}} 
If neither buy nor sell, return: 
{{"action_type" : "no"}} 
Please answer again.
\end{Verbatim}

\subsection{Post-Trading prompts}

\paragraph{FIRST\_DAY\_FINANCIAL\_REPORT}
\begin{Verbatim}[frame=single, breaklines=true, breaksymbol={}]
Company A has been listed for 10 years, deeply rooted in the chemical industry. However, the company's operations have encountered bottlenecks, with revenues declining over the past three years. In the short term, the stock price is expected to further decrease.
Although Company A's performance has declined over the past five years, the overall trend is stable. With the recent CEO change and the exploration of new business avenues, the new CEO appears more proactive compared to the previous one. The future operational outlook is expected to improve. Currently, Company A is at a low valuation, and I choose to buy into Company A. 

Company B, as a technology company, has just been listed for three years and is in a period of business growth. Last year, its revenue declined due to the overall tech environment, but the company's operations remain robust. According to the latest corporate news, it is expected that the future revenue growth rate will return to over 20%. In the short term, the stock price is expected to continue rising.
While Company B's operations are good, there is a history of concealing critical data before its IPO, casting doubt on the reliability of its revenue. I believe it is prudent to continue monitoring its performance in the next quarter before making investment decisions.
Company B recently received government inquiries regarding recent operational and stock price fluctuations, and it provided explanations while committing to allocate more resources to social services. I believe it will be challenging for it to expand into new businesses in the short term, and its stock price is likely to peak in the near term. 

The government recently held talks with both Company A and Company B, actively encouraging their contributions to society. Subsequently, agreements on government subsidies were signed with both companies. I believe the stock prices of these two companies should rise in the near term. Company A's performance has been declining over the past five years, and I believe it should be given a lower valuation. In the short term, I am bearish on Company A. On the other hand, Company B is at a relatively low valuation point, so I choose to buy into Company B and sell Company A.
\end{Verbatim}

\paragraph{SEASONAL\_FINANCIAL\_REPORT}
\begin{Verbatim}[frame=single, breaklines=true, breaksymbol={}]
Stock A: {stock_a_report} 
Stock B: {stock_b_report} 
\end{Verbatim}

\paragraph{POST\_MESSAGE\_PROMPT}
\begin{Verbatim}[frame=single, breaklines=true, breaksymbol={}]
The current trading day is over, please briefly post your trading tips on the forum and post them on the forum. What you post will be publicly visible to all traders. The responses contain only what needs to be posted.
\end{Verbatim}

\paragraph{NEXT\_DAY\_ESTIMATE\_PROMPT}
\begin{Verbatim}[frame=single, breaklines=true, breaksymbol={}]
based on the market information and forum information of the current trading day, please estimate whether you will buy and sell stock A and stock B tomorrow and whether you will choose a loan. 
Actions that are expected to take place are marked yes, and actions that will not take place are marked no. 
Return the result in JSON format, for example: 
{{"buy_A": "yes", "buy_B": "no", "sell_A": "yes", "sell_B": "no", "loan": "yes"}} 
\end{Verbatim}

\newpage

% \section{AI Agent Investors Trading and Behavior Data}
% \label{App:price}
% \subsection{Price statistics of the StockAgent's pending orders on Stock A and Stock B}
% \label{sec: reliability}

% \begin{figure}[ht]
%     \centering          
%     \includegraphics[width=1.0\textwidth]{img/trading_Gemini.pdf}
%     \caption{The Gemini-based agent order price.}
%     \label{fig:Gemini_price}
% \end{figure}
% \begin{figure}[ht]
%     \centering          
%     \includegraphics[width=1.0\textwidth]{img/trading_GPT.pdf}
%     \caption{The GPT-based agent order Price}
%     \label{fig:GPT_price}
% \end{figure}

\newpage

\section{Specific Experimental Data}
\label{App:data}

\begin{table}[ht]
\centering
\caption{The profit and loss of The StockAgent in different external environments. (Part I)}
\small
\vspace{-6pt}
\begin{tabular}{|c|r|r|r|r|}
\hline
\textbf{Agent} & \textbf{Non-finance} & \textbf{Non-BBS} & \textbf{Non-Loan} & \textbf{Non-statement} \\
\hline
0 & -156779.47 & -15745.19 & 529310.72 & -21934.03 \\
1 & -689016.46 & -4887.07 & 265022.24 & 13311.27 \\
2 & -223032.53 & 32037.27 & 302167.25 & 79.18 \\
3 & 96490.38 & -59496.88 & 612046.82 & -48529.67 \\
4 & -1745400.27 & -17709.74 & 537866.50 & 3881.38 \\
5 & -156246.98 & 11111.27 & 828530.87 & -42928.78 \\
6 & 41642.21 & -2767.66 & 1212495.30 & -2823.07 \\
7 & -1012926.31 & 12649.82 & 11560.30 & 14249.82 \\
8 & -496997.31 & 45591.91 & 529839.08 & 13353.14 \\
9 & 1614942.67 & -3375.10 & 32736.43 & 40877.82 \\
10 & 364317.30 & 44323.95 & 341684.56 & 20990.38 \\
11 & -858128.36 & -46191.30 & 580223.50 & 67595.85 \\
12 & 838929.98 & -56073.47 & 382722.50 & -13335.78 \\
13 & 797626.81 & -24126.24 & 771147.59 & -53765.72 \\
14 & 225132.36 & -18013.94 & 253366.11 & 35900.08 \\
15 & 272069.46 & 32821.59 & 323245.63 & 4679.30 \\
16 & -1411078.16 & 4113.27 & 442866.07 & 9863.41 \\
17 & -1578977.91 & 43809.94 & 80994.05 & 65456.97 \\
18 & 1801568.53 & 9090.56 & 474417.70 & -1770.68 \\
19 & -409366.23 & 24176.59 & 110635.39 & -14292.14 \\
20 & 905509.34 & 3929.84 & 10864.42 & -22021.30 \\
21 & 1577414.73 & -4757.95 & 698848.84 & 15979.12 \\
22 & -1640417.94 & -64931.81 & 135945.59 & -35297.88 \\
23 & 73159.00 & -54690.57 & 299238.95 & -37855.64 \\
24 & 748900.51 & -7049.34 & 386527.46 & 70325.96 \\
25 & -62019.05 & -25237.13 & 8280.01 & -24916.43 \\
26 & -300850.29 & 38222.78 & 171963.15 & 14488.07 \\
27 & -1958752.57 & -131992.39 & 1218441.76 & -106001.35 \\
28 & 1211683.67 & 4210.60 & 290570.31 & -22150.17 \\
29 & 45614.85 & -6475.64 & 848355.59 & 27480.70 \\
30 & 132912.09 & 35589.47 & 791398.93 & 40837.73 \\
31 & 236293.07 & 10999.13 & 159097.39 & -17798.89 \\
32 & 2848581.05 & 14230.15 & 78757.44 & 67849.57 \\
33 & 212978.91 & 9066.21 & -14833.09 & -1837.53 \\
34 & 778185.26 & 8579.54 & 740683.39 & -36053.55 \\
35 & 917514.35 & 16595.45 & 310930.50 & -16053.04 \\
36 & 848886.30 & -18358.99 & 709731.51 & 140262.95 \\
37 & -2513474.12 & -1030.03 & 113785.80 & -25919.84 \\
38 & 81247.80 & -7063.13 & 591216.43 & -81138.50 \\
39 & -3111160.94 & -32038.81 & 305957.45 & 140712.93 \\
40 & -1895940.05 & 4145.11 & 300702.02 & 176815.29 \\
41 & -1501258.97 & 33561.65 & 50588.31 & -117634.55 \\
42 & -589746.53 & 4824.77 & 249717.14 & -8286.80 \\
43 & -4077757.38 & 5916.18 & 40132.84 & -111623.81 \\
44 & 1761357.10 & -100695.62 & 572314.27 & -32365.55 \\
45 & 1807590.44 & -18840.46 & 157156.45 & 60180.71 \\
\hline
\end{tabular}
\label{tab:my_label-1}
\end{table}

\begin{table}[ht]
\centering
\caption{The profit and loss of The StockAgent in different external environments. (Part II)}
\small
\begin{tabular}{|c|r|r|r|r|}
\hline
\textbf{Agent} & \textbf{Non-finance} & \textbf{Non-BBS} & \textbf{Non-Loan} & \textbf{Non-statement} \\
\hline
46 & 12662.42 & 3772.76 & 383685.25 & -76783.70 \\
47 & -3063049.11 & 8833.90 & 392021.90 & 81286.26 \\
48 & 616907.76 & 58387.09 & 536951.70 & 72720.27 \\
49 & 353078.32 & -62948.08 & 318243.33 & -45226.56 \\
50 & 228931.33 & 20622.87 & 424742.21 & 33897.88 \\
51 & 719313.71 & -2728.61 & 491370.66 & 22523.10 \\
52 & -501576.75 & 22801.29 & 300612.63 & -87321.17 \\
53 & 342224.89 & -6181.24 & 255547.80 & 52706.61 \\
54 & 898350.85 & 27587.46 & 383961.94 & -75002.23 \\
55 & -5327194.97 & -21339.18 & 544499.77 & 10113.13 \\
56 & 488576.57 & -12792.26 & 722882.11 & -25783.49 \\
57 & -808982.90 & -19946.85 & 588549.75 & -1252.88 \\
58 & -1190697.04 & 10465.77 & 531740.66 & -101322.06 \\
59 & -500034.46 & -23961.59 & 413390.92 & 1211.81 \\
60 & 532080.40 & 5775.15 & 843165.02 & 42056.80 \\
61 & -2434682.31 & -115398.53 & 35161.81 & -149619.19 \\
62 & 1083612.64 & -1016.63 & 19242.98 & 228542.74 \\
63 & -461643.22 & 14305.04 & 163364.28 & -120267.07 \\
64 & -1424189.14 & -18740.44 & 92341.81 & -122033.91 \\
65 & -1739735.53 & -70111.68 & 294754.96 & -44432.65 \\
66 & 461148.72 & -70370.91 & 310600.75 & 201036.24 \\
67 & -2658874.18 & 7446.28 & 676927.29 & -198879.32 \\
68 & 1588560.26 & -14168.10 & 126915.69 & -9431.21 \\
69 & 282220.80 & 4018.11 & 336276.81 & 29871.23 \\
70 & -5981425.75 & 19946.07 & 272755.98 & -21129.84 \\
71 & 1160778.74 & -9981.79 & 361579.45 & 87998.81 \\
72 & 531484.39 & -25570.15 & 465131.78 & 13347.49 \\
73 & -3063651.24 & 12500.33 & 1214836.99 & -68289.60 \\
74 & 129353.09 & -33127.70 & 198820.09 & 111997.12 \\
75 & -5296080.05 & -9157.76 & 2606.87 & 48262.70 \\
76 & 451592.90 & -25893.98 & 140416.41 & -81306.28 \\
77 & 1207812.49 & 8187.97 & 735354.78 & 21750.48 \\
78 & -2720809.54 & 10195.86 & 244849.79 & -10004.15 \\
79 & 799885.57 & 2519.12 & 687269.08 & 1030.39 \\
80 & -3747958.85 & -61408.59 & 294798.79 & 19909.15 \\
81 & 137400.56 & 9212.03 & 162287.98 & -80712.20 \\
82 & 2959120.98 & -97978.93 & 44305.35 & 2847.82 \\
83 & -6252315.09 & 25597.62 & 303311.38 & 68417.17 \\
84 & -53447.31 & 34925.66 & 377790.09 & -29041.65 \\
85 & 912800.85 & -73578.94 & 688732.78 & 29587.13 \\
86 & -553463.36 & 11322.14 & 12502.56 & -94327.24 \\
87 & -1804927.38 & 21922.03 & -17648.21 & 148314.47 \\
88 & 1723276.30 & 12614.85 & 427244.26 & 114754.16 \\
89 & 1289542.65 & -25794.60 & 154190.44 & -190224.50 \\
90 & 177644.46 & 5402.24 & 282702.38 & -4732.47 \\

\hline
\end{tabular}

\label{tab:my_label-2}
\end{table}

\begin{table}[ht]
\centering
\caption{The profit and loss of The StockAgent in different external environments. (Part III)}
\small
\begin{tabular}{|c|r|r|r|r|}
\hline
\textbf{Agent} & \textbf{Non-finance} & \textbf{Non-BBS} & \textbf{Non-Loan} & \textbf{Non-statement} \\
\hline
91 & 934827.13 & -67237.83 & 849675.94 & 16116.05 \\
92 & 2861690.50 & -16324.46 & 746076.10 & -79070.65 \\
93 & 658919.86 & 48445.66 & 167233.71 & -26917.54 \\
94 & 2278357.75 & -2656.92 & 296735.39 & -16190.51 \\
95 & 313115.13 & -29362.15 & 397431.35 & 237527.72 \\
96 & -179647.13 & -24747.97 & 554465.32 & 112077.38 \\
97 & -411707.32 & -4148.34 & 9222.96 & -91673.99 \\
98 & -738854.23 & -47277.98 & 164133.79 & 3252.62 \\
99 & -823917.03 & 17103.78 & 524486.66 & 71102.30 \\
100 & 1584195.68 & 16727.10 & 320633.61 & -2694.53 \\
101 & -105625.81 & 13432.98 & 722484.19 & -214672.18 \\
102 & 1078835.44 & -42106.21 & 469175.82 & 142142.95 \\
103 & -416374.10 & 10042.74 & 1275400.81 & -38013.76 \\
104 & -100876.47 & 23999.53 & 526792.22 & -174117.30 \\
105 & 238619.32 & -33700.20 & 2935.75 & 57943.79 \\
106 & 1156134.16 & -36785.39 & 302372.82 & 115059.98 \\
107 & -1545273.59 & -30590.46 & 568349.53 & -48197.05 \\
108 & -158676.57 & -34301.66 & 385220.15 & -137224.76 \\
109 & -4832144.24 & -41565.12 & 754737.03 & -37944.45 \\
110 & 804477.14 & 19198.91 & 293116.45 & 154913.03 \\
111 & -2108163.37 & 19426.81 & 313628.42 & 98447.90 \\
112 & 1473247.28 & -18096.46 & 416890.50 & -5267.70 \\
113 & 17930.39 & -23490.68 & 97752.37 & -154982.44 \\
114 & 253412.35 & 9261.00 & 466398.35 & 2061.02 \\
115 & 760493.36 & -62972.02 & 108400.26 & -49677.36 \\
116 & -372384.36 & -16064.49 & 24059.16 & 10756.79 \\
117 & 436496.33 & 6573.43 & 698605.98 & 56100.39 \\
118 & 978512.74 & 129353.09 & -33127.70 & 198820.09 \\
119 & 97149.71 & 3096.97 & 317523.64 & -135140.62 \\
120 & 1131378.01 & -126979.89 & 409291.35 & -12781.11 \\
121 & 127356.38 & -55320.35 & 38528.35 & -58847.23 \\
122 & 319530.35 & 29578.85 & 165810.06 & 20526.14 \\
123 & 192615.38 & 15723.22 & 1221602.85 & -68048.40 \\
124 & 278460.73 & -36504.17 & 297191.56 & -3874.56 \\
125 & 1579135.23 & 9181.32 & 828106.74 & 93454.23 \\
126 & -220686.57 & -5590.81 & 796519.32 & 13694.02 \\
127 & 975456.71 & -2640.19 & 143708.72 & -4448.05 \\
128 & 837290.58 & 11494.98 & 83564.50 & -71064.33 \\
129 & 254980.31 & -3690.23 & -2033.85 & 31021.64 \\
130 & 168258.52 & 4997.85 & 728259.87 & -5645.30 \\
131 & -222640.20 & -112072.95 & 323166.74 & 18843.17 \\
132 & -404757.64 & 4146.77 & 667962.89 & -127583.66 \\
133 & 876309.64 & -47486.67 & 116975.48 & -51768.81 \\
134 & 260489.45 & 7213.84 & 604805.50 & 39809.62 \\
135 & 114797.03 & 4495.36 & 307748.76 & -27156.26 \\
\hline
\end{tabular}

\label{tab:my_label-3}
\end{table}

\begin{table}[ht]
\centering
\caption{The profit and loss of The StockAgent in different external environments. (Part IV)}
\small
\begin{tabular}{|c|r|r|r|r|}
\hline
\textbf{Agent} & \textbf{Non-finance} & \textbf{Non-BBS} & \textbf{Non-Loan} & \textbf{Non-statement} \\
\hline

136 & 158071.54 & -29013.38 & 339010.88 & 55349.71 \\
137 & -1634393.68 & 30977.69 & 8567.59 & -30253.51 \\
138 & -818855.66 & 30280.59 & 252178.08 & 87074.52 \\
139 & -1982207.87 & -47205.84 & 19447.99 & 39624.16 \\
140 & 716200.30 & 10153.13 & 520477.17 & -56959.28 \\
141 & -1316159.14 & -25600.88 & 201578.75 & -75984.20 \\
142 & -2122498.47 & -32407.44 & 362434.52 & 68708.69 \\
143 & -81005.19 & -27980.01 & 408372.65 & -46057.02 \\
144 & 1528723.84 & 8680.60 & 583657.01 & -38296.30 \\
145 & 48922.99 & -19171.60 & 264185.31 & 8348.06 \\
146 & -3500332.09 & 48582.49 & 417507.69 & 79731.14 \\
147 & -566369.24 & 22291.64 & 480302.25 & 18142.74 \\
148 & 937040.60 & 2338.45 & 302603.11 & 117842.83 \\
149 & -1203052.08 & 10293.69 & 305862.03 & -11244.60 \\
150 & -833876.79 & -19278.37 & 404933.97 & -117929.06 \\
151 & -1201365.32 & -102420.58 & 544365.82 & -122178.47 \\
152 & -162150.90 & -70851.31 & 730527.64 & -78328.22 \\
153 & 406189.11 & 24523.49 & 594586.15 & 37465.05 \\
154 & -7362732.30 & 38329.20 & 536580.18 & 77726.95 \\
155 & -636852.37 & 10226.80 & 423032.52 & 76374.19 \\
156 & 202177.52 & 29632.82 & 826126.95 & 57099.00 \\
157 & 566202.41 & 2891.15 & -22716.95 & -18899.30 \\
158 & 266419.81 & 1969.49 & 10882.61 & 73107.16 \\
159 & -2012979.15 & -68800.46 & 180350.24 & -139283.37 \\
160 & 241126.68 & 22405.01 & 94751.05 & 42665.07 \\
161 & -561857.69 & -90375.20 & 340652.11 & 67409.74 \\
162 & 1306954.30 & -26049.80 & 308934.18 & -28489.11 \\
163 & -383936.14 & 29688.27 & 668817.71 & -76571.81 \\
164 & 458060.67 & -14624.57 & 122174.90 & -25700.32 \\
165 & 1164634.00 & 7482.82 & 309735.64 & 21877.42 \\
166 & 341680.69 & 9165.37 & 274799.07 & -44899.44 \\
167 & 305821.63 & 4002.65 & 370692.10 & 155314.38 \\
168 & -975037.64 & -31341.71 & 482347.41 & -63424.57 \\
169 & -86638.87 & 33707.39 & 1243389.98 & 31420.52 \\
170 & -1910620.53 & 12348.98 & 165727.12 & -79898.49 \\
171 & 839119.86 & 16899.22 & 3224.07 & 116446.44 \\
172 & 1199968.82 & -103766.17 & 134233.48 & 9164.44 \\
173 & -940358.54 & -42810.10 & 728935.27 & -36908.81 \\
174 & -712389.21 & -48197.88 & 289285.14 & -20553.23 \\
175 & 456228.62 & -345.93 & 665963.86 & -46564.95 \\
176 & 1036506.06 & -39655.87 & 301304.01 & -19053.75 \\
177 & -218618.93 & -51277.02 & 125480.92 & 87788.29 \\
178 & 606788.54 & -17325.39 & 3394.30 & 23747.28 \\
179 & -2198442.90 & -40189.44 & 302812.03 & -807.85 \\
180 & 1640630.27 & 29392.07 & 362514.94 & -106331.21 \\

\hline
\end{tabular}

\label{tab:my_label-4}
\end{table}

\begin{table}[ht]
\centering
\caption{The profit and loss of The StockAgent in different external environments. (Part V)}
\small
\begin{tabular}{|c|r|r|r|r|}
\hline
\textbf{Agent} & \textbf{Non-finance} & \textbf{Non-BBS} & \textbf{Non-Loan} & \textbf{Non-statement} \\
\hline
181 & 2317537.84 & 5808.46 & 721660.34 & 148478.58 \\
182 & -403863.94 & -13871.31 & 41009.84 & 13326.86 \\
183 & 394682.37 & -10578.68 & -2178.92 & 50915.96 \\
184 & 75517.66 & -2780.68 & 409020.55 & -101668.46 \\
185 & -1331009.66 & -22814.79 & 175132.65 & 53716.30 \\
186 & -1906789.51 & -10820.08 & 273817.38 & -15703.72 \\
187 & 172075.45 & 21415.58 & 843905.59 & 48538.56 \\
188 & 1157491.52 & 2684.72 & 758201.99 & -52973.95 \\
189 & -1088761.91 & 323.17 & 132951.63 & 39117.55 \\
190 & 928545.17 & 5150.54 & 318504.07 & 29203.14 \\
191 & -1512313.11 & -28458.68 & 387082.08 & -5321.91 \\
192 & 959083.41 & 16700.97 & 572748.00 & -54938.08 \\
193 & -1471037.95 & 35837.79 & 41611.86 & -66058.87 \\
194 & -302136.97 & 15665.27 & 162051.55 & 24968.25 \\
195 & 2210228.93 & 7161.37 & 538696.16 & 148478.58 \\
196 & -672293.47 & 389.16 & 301130.48 & 13326.86 \\
197 & 400612.61 & -85830.30 & 720185.25 & 50915.96 \\
198 & 659949.64 & -34345.24 & 465646.57 & -101668.46 \\
199 & -3347763.52 & 31579.47 & 1258219.66 & 48538.56 \\
\hline
\end{tabular}

\label{tab:my_label-5}
\end{table}

\begin{table}[ht]
\centering
\caption{StockAgent's stock price movement in the case of ablating the external environment (Part I)}
\small
\label{tab:partial_profitability_part1}
\begin{tabular}{|c|cc|cc|cc|}
\hline
\textbf{Round} & \textbf{No\_Info\_A} & \textbf{No\_Info\_B} & \textbf{NoBBS\_A} & \textbf{NoBBS\_B} & \textbf{No\_State\_A} & \textbf{No\_State\_B} \\
\hline
1 & 29.0 & 39.0 & 28.0 & 39.1 & 30.2 & 44.4 \\
2 & 29.0 & 39.1 & 28.9 & 38.0 & 29.7 & 44.3 \\
3 & 28.5 & 39.1 & 28.5 & 38.0 & 29.7 & 44.2 \\
4 & 29.0 & 39.1 & 27.8 & 37.7 & 29.7 & 44.2 \\
5 & 29.0 & 37.4 & 27.5 & 37.8 & 29.2 & 44.2 \\
6 & 28.5 & 37.4 & 27.7 & 38.0 & 29.4 & 44.2 \\
7 & 28.3 & 37.2 & 27.8 & 37.8 & 29.1 & 44.0 \\
8 & 27.9 & 37.2 & 27.1 & 38.2 & 29.2 & 43.8 \\
9 & 28.1 & 37.2 & 28.0 & 38.0 & 29.1 & 43.0 \\
10 & 27.8 & 36.5 & 28.2 & 39.3 & 29.1 & 43.0 \\
11 & 28.0 & 36.4 & 28.1 & 39.2 & 29.0 & 43.0 \\
12 & 27.0 & 36.2 & 28.0 & 39.2 & 29.0 & 43.0 \\
13 & 27.0 & 35.9 & 28.6 & 39.0 & 29.0 & 42.1 \\
14 & 26.3 & 35.1 & 28.1 & 39.2 & 29.5 & 42.2 \\
15 & 26.4 & 35.1 & 28.3 & 39.4 & 28.9 & 42.1 \\
16 & 26.3 & 34.9 & 28.4 & 39.1 & 28.9 & 42.4 \\
17 & 26.4 & 34.8 & 28.2 & 39.1 & 28.5 & 42.3 \\
18 & 26.1 & 34.8 & 28.5 & 39.1 & 28.5 & 41.9 \\
19 & 26.3 & 34.5 & 28.9 & 39.0 & 28.5 & 41.3 \\
20 & 26.3 & 34.5 & 28.3 & 39.0 & 28.5 & 42.0 \\
21 & 26.3 & 34.2 & 28.7 & 39.0 & 28.8 & 42.0 \\
22 & 26.1 & 34.2 & 28.5 & 38.9 & 28.5 & 42.0 \\
23 & 26.1 & 34.0 & 28.5 & 38.3 & 28.5 & 33.0 \\
24 & 26.0 & 34.2 & 27.6 & 38.7 & 28.2 & 33.0 \\
25 & 25.9 & 34.1 & 27.9 & 38.5 & 28.0 & 33.0 \\
26 & 25.9 & 34.0 & 27.4 & 38.6 & 28.0 & 33.3 \\
27 & 25.8 & 33.9 & 27.1 & 38.4 & 28.1 & 33.1 \\
28 & 25.3 & 33.5 & 26.0 & 38.0 & 28.1 & 33.1 \\
29 & 25.3 & 33.4 & 26.2 & 37.7 & 28.1 & 33.1 \\
30 & 25.3 & 33.4 & 26.4 & 37.8 & 28.0 & 33.1 \\ \hline
\end{tabular}
\end{table}

\begin{table}[ht]
\centering
\caption{StockAgent's stock price movement in the case of ablating the external environment (Part II)}
\label{tab:partial_profitability_part2}
\small
\begin{tabular}{|c|cc|cc|}
\hline
\textbf{Round} & \textbf{No\_Loan\_A} & \textbf{No\_Loan\_B} & \textbf{No\_Interest\_Change\_A} & \textbf{No\_Interest\_Change\_B} \\
\hline
1 & 29.0 & 39.1 & 30.0 & 40.0 \\
2 & 29.9 & 38.0 & 30.0 & 40.7 \\
3 & 29.5 & 38.0 & 30.0 & 40.7 \\
4 & 28.8 & 37.7 & 30.0 & 40.7 \\
5 & 28.5 & 37.8 & 30.0 & 40.7 \\
6 & 28.7 & 38.0 & 33.0 & 40.7 \\
7 & 28.8 & 37.8 & 33.0 & 40.7 \\
8 & 28.1 & 38.2 & 33.0 & 40.7 \\
9 & 28.0 & 38.0 & 35.5 & 40.7 \\
10 & 28.2 & 39.3 & 35.5 & 40.7 \\
11 & 28.1 & 39.2 & 35.5 & 40.7 \\
12 & 28.0 & 39.2 & 35.5 & 40.7 \\
13 & 28.6 & 39.0 & 35.5 & 40.7 \\
14 & 28.1 & 39.2 & 36.7 & 41.5 \\
15 & 28.3 & 39.4 & 36.7 & 41.6 \\
16 & 28.4 & 39.1 & 41.8 & 41.6 \\
17 & 28.2 & 39.1 & 42.7 & 41.8 \\
18 & 28.1 & 39.1 & 43.0 & 41.8 \\
19 & 27.9 & 39.0 & 43.0 & 41.8 \\
20 & 28.1 & 39.0 & 43.0 & 41.8 \\
21 & 28.1 & 39.0 & 43.0 & 43.0 \\
22 & 26.5 & 38.9 & 46.8 & 43.0 \\
23 & 26.6 & 38.3 & 47.5 & 43.0 \\
24 & 27.6 & 38.7 & 47.5 & 45.5 \\
25 & 27.9 & 38.5 & 50.5 & 45.5 \\
26 & 27.4 & 38.6 & 51.8 & 45.5 \\
27 & 27.1 & 38.4 & 52.0 & 45.5 \\
28 & 26.0 & 38.0 & 56.9 & 45.5 \\
29 & 26.2 & 37.7 & 56.9 & 45.5 \\
30 & 26.4 & 37.8 & 62.5 & 46.0 \\
\hline
\end{tabular}
\end{table}

\end{document}